\documentclass[letterpaper,prd,aps,nofootinbib]{revtex4}
\usepackage{amssymb}
\usepackage{bm}
\usepackage{enumerate}
\usepackage{amssymb}
\usepackage{amsmath}
\usepackage{feynmf}
\usepackage{slashed}
\usepackage{wrapfig}
\usepackage[caption=false]{subfig}
\usepackage{graphicx}
\usepackage{filecontents}
\usepackage{mathtools}
\usepackage{xcolor}

\newcommand\be{\begin{equation}}
\newcommand\ee{\end{equation}}
\begin{document}

\title{\Large \bf Early Matter Domination from Long-Lived Particles in the Visible Sector} 

\author{Rouzbeh Allahverdi$^{1}$}
\author{Jacek K.~Osi{\'n}ski$^{2}$}
\affiliation{$^{1}$~Department of Physics and Astronomy, University of New Mexico, Albuquerque, NM 87131, USA}
\affiliation{$^{2}$~Astrocent, Nicolaus Copernicus Astronomical Center of the Polish Academy of Sciences, ul.~Rektorska 4, 00-614 Warsaw, Poland}

\begin{abstract}

We show that a nonstandard cosmological history with a period of early matter domination driven by a sub-TeV visible-sector particle can arise rather naturally. This scenario involves a long-lived standard model singlet that acquires a thermal abundance at high temperatures from decays and inverse decays of a parent particle with SM charge(s), and subsequently dominates the energy density of the Universe as a frozen species. Entropy generation at the end of early matter domination dilutes the abundance of dangerous relics (such as gravitinos) by a factor as large as $10^4$. The scenario can accommodate the correct dark matter relic abundance for cases with $\langle \sigma_{\rm ann} v \rangle_{\rm f} \lessgtr 3 \times 10^{-26}$cm$^3$s$^{-1}$. 
More importantly, the allowed parameter space can be directly probed by proposed searches for neutral long-lived particles at the energy frontier, allowing us to use particle physics experiments to reconstruct the cosmological history just prior to big bang nucleosynthesis. 

\end{abstract}
\maketitle

\section{Introduction}

Despite various lines of evidence for the existence of dark matter (DM)~\cite{BHS}, its identity remains a major problem at the interface of cosmology and particle physics. The relic abundance of DM depends on its particle physics origin as well as the thermal history of the early Universe. Thermal freeze-out in a radiation-dominated (RD) Universe can explain the DM content of the Universe if the thermally averaged annihilation rate takes the nominal value $\langle \sigma_{\rm ann} v \rangle_{\rm f} = 3 \times 10^{-26}$cm$^3$s$^{-1}$ at the time of freeze-out, when \(T = T_{\rm f}\). However, the correct relic abundance can also be obtained for much larger or smaller values of $\langle \sigma_{\rm ann} v \rangle_{\rm f}$ if the Universe is not in a RD phase at the time of freeze-out~\cite{KT}. An epoch of early matter domination (EMD), which is a generic feature of early Universe models from string theory constructions~\cite{KSW}, provides an important such example. EMD is driven by a matter-like species that comes to dominate the energy density of the Universe and decays to establish RD prior to big bang nucleosynthesis (BBN). Various production mechanisms during EMD can yield the correct DM abundance for both $\langle \sigma_{\rm ann} v \rangle_{\rm f} \lessgtr 3 \times 10^{-26}$cm$^3$s$^{-1}$.

A component whose equation of state is the same as matter can lead to an EMD phase in the postinflationary Universe if it constitutes a sizeable fraction of the total energy density and is sufficiently long lived. In the context of string theory, this can arise from coherent oscillations of string moduli that are displaced from the minimum of their potential during inflation. Modulus fields have long lifetimes due to their gravitationally suppressed couplings to other fields. An EMD era may also be driven by long-lived nonrelativistic quanta that are produced in the postinflationary Universe and dominate the energy density before decaying. This can happen, for example, in models that involve hidden sectors~\cite{Ng,Hooper,Scott2,Cirelli}. Scenarios with EMD have interesting observable predictions that can be tested via cosmological observations and DM indirect detection searches~\cite{E1,E2,E3,E4,E5}. A natural question that arises is whether one could also directly probe the particle(s) driving an epoch of EMD in laboratory experiments.      

Motivated by this, we present a scenario where a visible-sector particle with a mass around the weak scale drives an EMD era. This scenario involves a standard model (SM) singlet, $N$, that reaches thermal equilibrium in a RD Universe at temperatures much higher than its mass as a result of decays and inverse decays of a parent particle, $X$, with SM charges. $N$ quanta 
then maintain a frozen comoving number density, dominate the energy density of the Universe, and eventually decay to SM particles via an effective interaction mediated by $X$. Decay of $N$ is a higher-order process that is also suppressed by three-body phase space and by powers of $m_N/m_X$, which results in a sufficiently long lifetime that can go all the way to the onset of BBN, with $\tau_N \sim 0.1$ s. The late decay of $N$ dilutes dangerous relics like unstable gravitinos, as well as any overabundance of DM or baryon asymmetry, from earlier stages by a factor as large as $10^4$. Depending on the relation between $m_N$ and the DM mass,
the observed DM abundance may be obtained for both large and small values of $\langle \sigma_{\rm ann} v \rangle_{\rm f}$. Moreover, the parameter space of this scenario may be directly probed at the energy frontier, such as by the proposed searches for long-lived particles at the Large Hadron Collider (LHC)~\cite{MAT1,MAT2,MAT3}.               

The rest of this paper is organized as follows. In Section II, we discuss our scenario and the conditions that need to be satisfied for its success. We also present a minimal extension of the SM that can explicitly realize the scenario. In Section III, we present our main results and identify the allowed regions of the parameter space of the scenario. We focus on the $m_N-\tau_N$ plane in Section IV and determine regions that yield the correct DM abundance. We also discuss the prospects for probing the $m_N-\tau_N$ plane in tandem with the long-lived particle searches at the LHC. We conclude the paper
in Section V. Some details of our calculations are presented in the Appendix.  

\section{Early Matter Domination from the Visible Sector}

Our scenario involves a SM singlet in the visible sector, $N$, that reaches equilibrium at early times via interactions with the thermal bath. It behaves as a matter component after becoming nonrelativistic and, provided that it is sufficiently long lived, dominates the energy density of the Universe. This results in an epoch of EMD that ends when $N$ decay establishes a RD Universe and sets the stage for BBN.

\subsection{The Scenario}

For a quantitative study of this scenario, let us consider an extension of the SM with the following Lagrangian:
\begin{eqnarray} \label{Lnew}
&& {\cal L} = {\cal L}_{\rm SM} + {\cal L}_{\rm new} \, , \nonumber \\
&& {\cal L}_{\rm new} \supset h X N \psi + h^{\prime} X^{\dagger} \psi \psi + {\rm h.c.} .
\end{eqnarray}
Here $\psi$ collectively denotes the SM fermions, $X$ is a scalar with mass $m_X$ that is charged under the SM, and $N$ is a Majorana fermion with mass $m_N \ll m_X$ that is a SM singlet. Since the SM is a chiral theory, gauge invariance implies that each $\psi$ in Eq.~(\ref{Lnew}) has a certain chirality. At energies $E \ll m_X$, which is relevant for $N$ decay, the above Lagrangian gives rise to an effective four-fermion interaction $h h^{\prime \dagger} N \psi \psi \psi/m^2_X$\footnote{Examples of gauge-invariant effective interactions, in terms of left-handed (LH) chiral fermions, include $N u^c u^c d^c$, $N L L e^c$, and $N Q L d^c$. These arise when $X$ is a color triplet, an electroweak doublet, and a scalar leptoquark respectively. We will discuss the first case in detail shortly.}. This results in $N$ decay into three-body final states consisting of the SM fermions. 

Starting in a RD phase at $T \gtrsim m_X$, established at the end of inflationary reheating (for reviews, see~\cite{ABCM,Aminreview}) or from the decay of a heavy modulus, the important stages of the thermal history in our scenario, arranged by decreasing Hubble rate, are as follows:
\vskip 2mm
\noindent
{\bf (1)} $H \gtrsim H(T = m_X)$. At this stage, $X$ is in thermal equilibrium due to its gauge interactions with the SM particles. $N$ also acquires a thermal abundance, $n_N \propto T^3$ and $\rho_N \propto T^4$, provided that the partial decay width of $X$ to $N$ satisfies:
\begin{equation} \label{equi}
\Gamma_{X \rightarrow N} \gtrsim H(T = m_X) . 
\end{equation}
This ensures that $X$ decay and inverse decay reach equilibrium before $X$ becomes nonrelativistic (see Appendix A for details).
\vskip 2mm
\noindent
{\bf (2)} $H(T = m_X) > H \gtrsim H(T = m_N)$. During this stage, $N$ particles are relativistic and $n_N \propto a^{-3}$ (where $a$ is the scale factor of the Universe). More precisely, $g_* n_N \propto a^{-3}$ as long as $N$ is in kinetic equilibrium with the thermal bath, while $n_N \propto a^{-3}$ when it is kinetically decoupled. Note, however, that the difference is negligible since the number of relativistic degrees of freedom, $g_*$, changes minimally (if at all) for $m_N \lesssim T < m_X$.
\vskip 2mm
\noindent
{\bf (3)} $H(T = m_N) > H \gtrsim H_{\rm dom}$. $N$ quanta become nonrelativistic during this stage. Their comoving energy density becomes frozen and remains constant provided that the rate for $N N \rightarrow \psi \psi$ and $N \psi \rightarrow \psi \psi$ satisfy: 
\begin{eqnarray}
&& \Gamma_{N N \rightarrow \psi \psi^*} < H(T = m_N) \label{selfann} \, , \\
&& \Gamma_{N \psi \rightarrow \psi^* \psi^*} < H(T= m_N) \label{ann} \, .
\end{eqnarray}
These relations ensure that $N$ self-annihilation and its annihilation with SM particles are inefficient at $T \lesssim m_N$.

Once the energy density of $N$ becomes comparable to that of radiation, $\rho_N \simeq \rho_{\rm R}$, $N$ starts to dominate the energy density of the Universe. This happens at $H = H_{\rm dom}$, which can be found from the following: 
\begin{equation} \label{Hdom1}
{2 \pi^2 \over 30} m^4_N \left({H_{\rm dom} \over H(T = m_N)}\right)^{3/2} \simeq {g_{*N} \pi^2 \over 30} m^4_N \left({H_{\rm dom} \over H(T = m_N)}\right)^{2} \left({g_{*N} \over g_{*{\rm dom}}}\right)^{1/3} .
\end{equation}
We have used the fact that $\rho_N \propto a^{-3}$, while $\rho_{\rm R} \propto g^{-1/3}_* a^{-4}$, and $a(t) \propto H^{-1/2}$ during RD\footnote{Recall that the comoving entropy density of the Universe is constant during RD. This implies that $g_* T^3 \propto a^{-3}$, and hence $\rho_{\rm R} \propto g_* T^4 \propto g^{-1/3}_* a^{-4}$.}. The number of relativistic degrees of freedom at $T = m_N$ and $T = T_{\rm dom}$ are given by $g_{*N}$ and $g_{*{\rm dom}}$ respectively. The factor of 2 on the LH side of Eq.~(\ref{Hdom1}) accounts for the two degrees of freedom associated with the Majorana fermion $N$. Therefore, we arrive at:
\begin{equation} \label{Hdom2}
H_{\rm dom} \simeq {4 g^{2/3}_{*{\rm dom}} \over g^{8/3}_{*N}} H(T = m_N). 
\end{equation}
The necessary condition for $N$ dominance is:
\begin{equation} \label{dom}
\Gamma_N < H_{\rm dom},
\end{equation}
where $\Gamma_N$ is the width of $N$ decay into three SM particles.
\vskip 2mm
\noindent
{\bf (4)} $H_{\rm dom} > H \gtrsim \Gamma_N$. The Universe is in an EMD era driven by $N$ during this stage, which ends when $N$ quanta decay. 
$N$ decay reheats the Universe to a temperature $T_{\rm dec}$. Note that $N$ decay must complete before the onset of BBN, and hence $T_{\rm dec} > T_{\rm BBN} \simeq 3$ MeV\footnote{An \(\mathcal{O}(1)\) MeV lower bound on the final reheat temperature is studied in \cite{Kohri} considering the thermalization process for neutrinos including neutrino oscillations and self interactions for both radiative and hadronic decays.}, which implies that:
\begin{equation} \label{onset}
\Gamma_N \gtrsim H_{\rm BBN} \sim 10 ~ {\rm s}^{-1}.
\end{equation}
The decay releases entropy and dilutes any preexisting relic abundance by the following factor:
\begin{equation} \label{d1}
d \equiv {s_{\rm after} \over s_{\rm before}}.
\end{equation}
Here, $s_{\rm before}$ is the entropy density of radiation existing from prior stages at $H = \Gamma_N$, while $s_{\rm after}$ is the entropy density generated by $N$ decay. They are given by:
\begin{eqnarray} \label{d2}
&& s_{\rm before} = {2 \pi^2 \over 45} g_{*{\rm dom}} T^3_{\rm dom} \left({\Gamma_N \over H_{\rm dom}}\right)^2 \, , \nonumber \\
&& s_{\rm after} = {2 \pi^2 \over 45} g_{*{\rm dec}} T^3_{\rm dec} \, ,
\end{eqnarray}
where we have used the fact that $a(t) \propto H^{-2/3}$ during the EMD epoch. The Universe enters a RD phase at $H = \Gamma_N$, and hence $T_{\rm dec} \simeq (90/\pi^2 g_{*{\rm dec}})^{1/4} (\Gamma_N M_{\rm P})^{1/2}$. We therefore find:
\begin{equation} \label{d3}
d \simeq \left({g_{*{\rm dec}} \over g_{*\rm dom}} \right)^{1/4} \left({H_{\rm dom} \over \Gamma_N}\right)^{1/2}.
\end{equation}

\vspace{-0.21cm}
\subsection{An Explicit Realization}
\vspace{-0.14cm}

We now present a specific realization of the scenario introduced above. It is based on a minimal extension of the SM that was proposed for low-scale baryogenesis and DM~\cite{RB} (see~\cite{Rabi} for a supersymmetric version). The Lagrangian (using two-component Weyl fermions) is:
\begin{eqnarray} \label{lagran}
{\cal L} \supset (h_{i} X N u^c_{i} + h^{\prime}_{i j} X^* {d^c_i} d^c_j + {1 \over 2} m_N N N + {\rm h.c.}) + 
m^2_X |X|^2 .
\end{eqnarray}
Here, $u^c$ and $d^c$ stand for the LH up-type and down-type antiquarks respectively. $i$ and $j$ are flavor indices, color indices are omitted for simplicity, and $h^\prime_{i j}$ is an antisymmetric tensor. $X$ is an iso-singlet color-triplet scalar of hypercharge $+4/3$. $N$ is a singlet fermion, with $m_N \ll m_X$, which may be charged under a higher-ranked gauge group that includes the SM.

At energies $E \ll m_X$, one finds an effective four-fermion interaction $N u^c_i d^c_j d^c_k$ after integrating out $X$. This results in $N$ decay to three quark, and since $N$ is a Majorana fermion, three antiquark final states. $N$ decay can be the origin of baryon asymmetry in the model as described in~\cite{Rabi}. Moreover, this model can explain the DM content of the Universe. In the supersymmetric version, the scalar partner of $N$ is a natural DM candidate~\cite{Rabi,ADMS}. In the nonsupersymmetric version, a second copy of $N$ that has approximately the same mass as the proton is a viable DM candidate~\cite{RB}. This can in addition address the baryon-DM coincidence puzzle.

This model also has testable predictions for phenomenology. 
The baryon-number-violating couplings of $X$ result in  $n-{\bar n}$ oscillation. It also has novel monojet/monotop signals, as well as dijet events, at the LHC~\cite{LHC1,LHC2,ADD}. The most stringent experimental bound on the model parameters comes from double proton decay $p p \rightarrow K^+ K^+$. For $m_X \sim 10$ TeV and $m_N \sim 1$ TeV, so that the model will be testable at colliders, the resulting limit is $\vert h_1 h^{\prime}_{12} \vert < 10^{-4}$~\cite{DM}. There are also bounds from $K^0_s$-${\bar K}^0_s$ and $B^0_s$-${\bar B}^0_s$ mixing, but they can be easily satisfied in the allowed parameter space.

\section{Results}

In this section we present our results. First, we would like to demonstrate that all of the conditions needed for the success of our scenario, given in Eqs.~(\ref{equi},\ref{selfann},\ref{ann},\ref{dom},\ref{onset}), can indeed be simultaneously satisfied. Note that, as explained before, $\psi$ particles are SM fermions of a specific chirality. For simplicity, we consider $X$ couplings to given flavor combinations in $N \psi$ and $\psi \psi$. The first term in the Lagrangian in Eq.~(\ref{Lnew}) results in the following (rest frame) partial decay width for $X$ (provided that $m_N \ll m_X$):
\begin{equation} \label{Xdec}
\Gamma_{X \rightarrow N} \simeq {\vert h \vert^2 \over 16 \pi} m_X.  
\end{equation}
The same term also leads to the following rate for $N$ self-annihilation at energies $E \ll m_X$:
\begin{equation} \label{Nself}
\Gamma_{N N \rightarrow \psi \psi^*} \simeq C_1 {\vert h \vert^4 \over 16 \pi} {E^2 \over m^4_X} n_N .    
\end{equation}
Here, $C_1$ is a multiplicity factor that is associated with gauge charges of the final state $\psi \psi^*$. 

Combination of the two terms in Eq.~(\ref{Lnew}) allows $N$ annihilation with $\psi$ at the following rate (at energies below $m_X$):
\begin{equation} \label{Nann}
\Gamma_{N \psi \rightarrow \psi^* \psi^*} \simeq 3 C_2 {\vert h \vert^2 \vert h^{\prime} \vert^2 \over 16 \pi} {E^2 \over m^4_X} n_\psi ,
\end{equation}
where $C_2$ is a multiplicity factor that counts all possible combinations of gauge charges in the initial and final states. The factor of $3$ on the RH side is due to the fact that $N$ can annihilate with any of the three $\psi$'s that show up in the effective interaction $N \psi \psi \psi$. For the number densities in Eqs.~(\ref{Nself},\ref{Nann}) we use the equilibrium value for a relativistic fermionic degree of freedom $3 \zeta(3) T^3 / 4 \pi^2$. Finally, the width for three-body decay of $N$ in its rest frame is:
\begin{equation} \label{Ndec}
\Gamma_N = 2 C_2 {\vert h \vert^2 \vert h^{\prime} \vert^2 \over 128 \cdot 192 \pi^3} {m^5_N \over m^4_X},
\end{equation}
where the factor of 2 on the RH side arises because $N$, which is a Majorana fermion, decays to both $\psi \psi \psi$ and $\psi^* \psi^* \psi^*$ final states.

To put things in perspective, consider the model described by Eq.~(\ref{lagran}). For simplicity, we consider $X$ coupling to a single flavor combination $u^c_i d^c_j d^c_k$ where $i,~j,~k$ are fixed (with $j \neq k$). In this model, $X \rightarrow N u_i$ decay brings $N$ into equilibrium at $T \gtrsim m_X$. The multiplicity factor for $N$ self-annihilation in Eq.~(\ref{Nself}) is $C_1 = 3$ because of the color of $u_i$ in the final state. The multiplicity factor for $N$ annihilation Eq.~(\ref{Nann}) and $N$ decay Eq.~(\ref{Ndec}) is $C_2 = 6$ due to possible color combinations in $u_i d_j d_k$. In general, $C_1$ and $C_2$ are model-dependent factors that can be found for a given model. Also, we need to sum over all flavor combinations of $\psi$'s that contribute to the various rates in Eqs.~(\ref{Xdec}--\ref{Ndec}).

\begin{figure}[ht!]
    \centering
    \includegraphics[width=0.49\textwidth, trim = .6cm 0cm 1cm 0cm, clip = true]{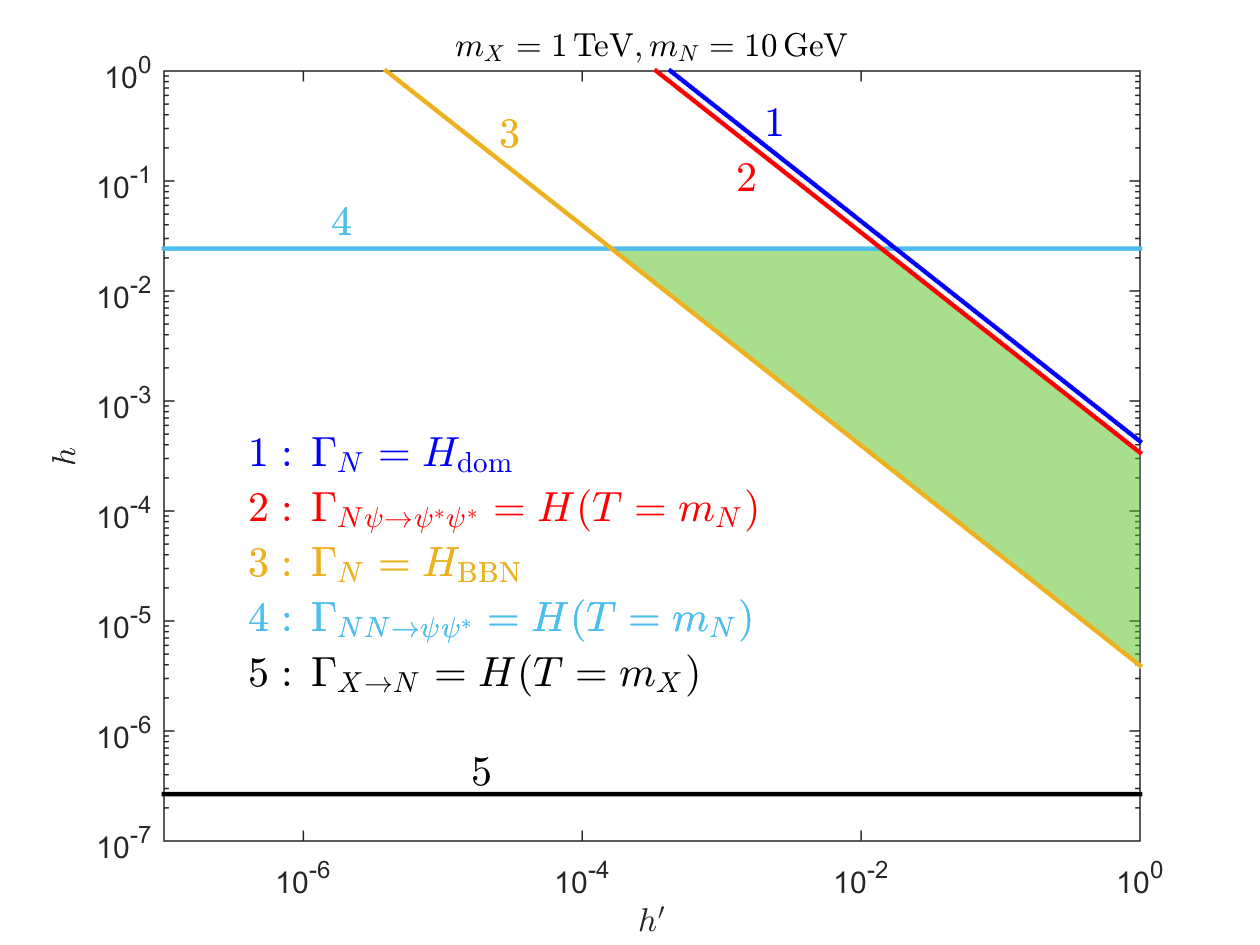}
    \includegraphics[width=0.49\textwidth, trim = .6cm 0cm 1cm 0cm, clip = true]{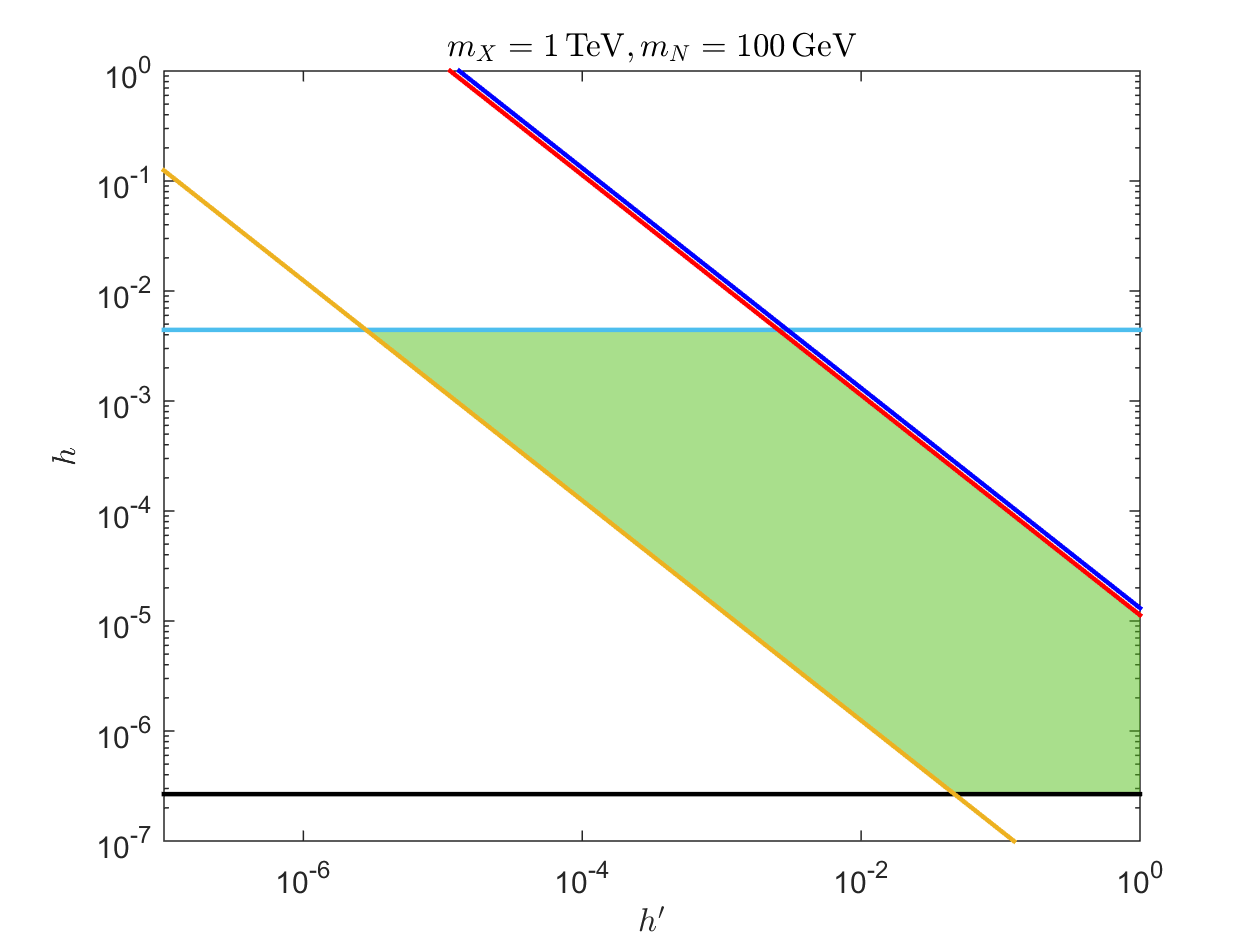}
    \includegraphics[width=0.49\textwidth, trim = .6cm 0cm 1cm 0cm, clip = true]{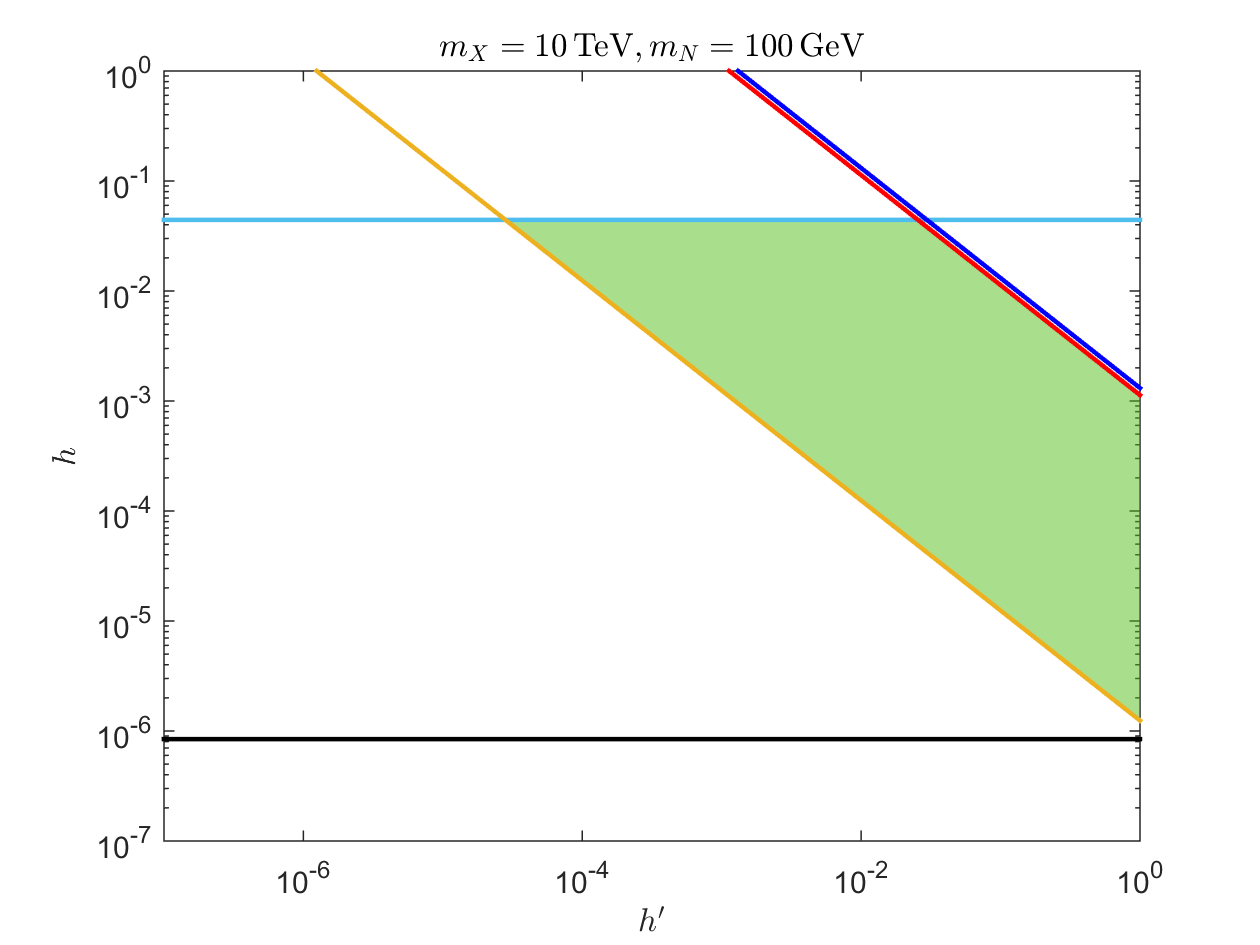}
    \includegraphics[width=0.49\textwidth, trim = .6cm 0cm 1cm 0cm, clip = true]{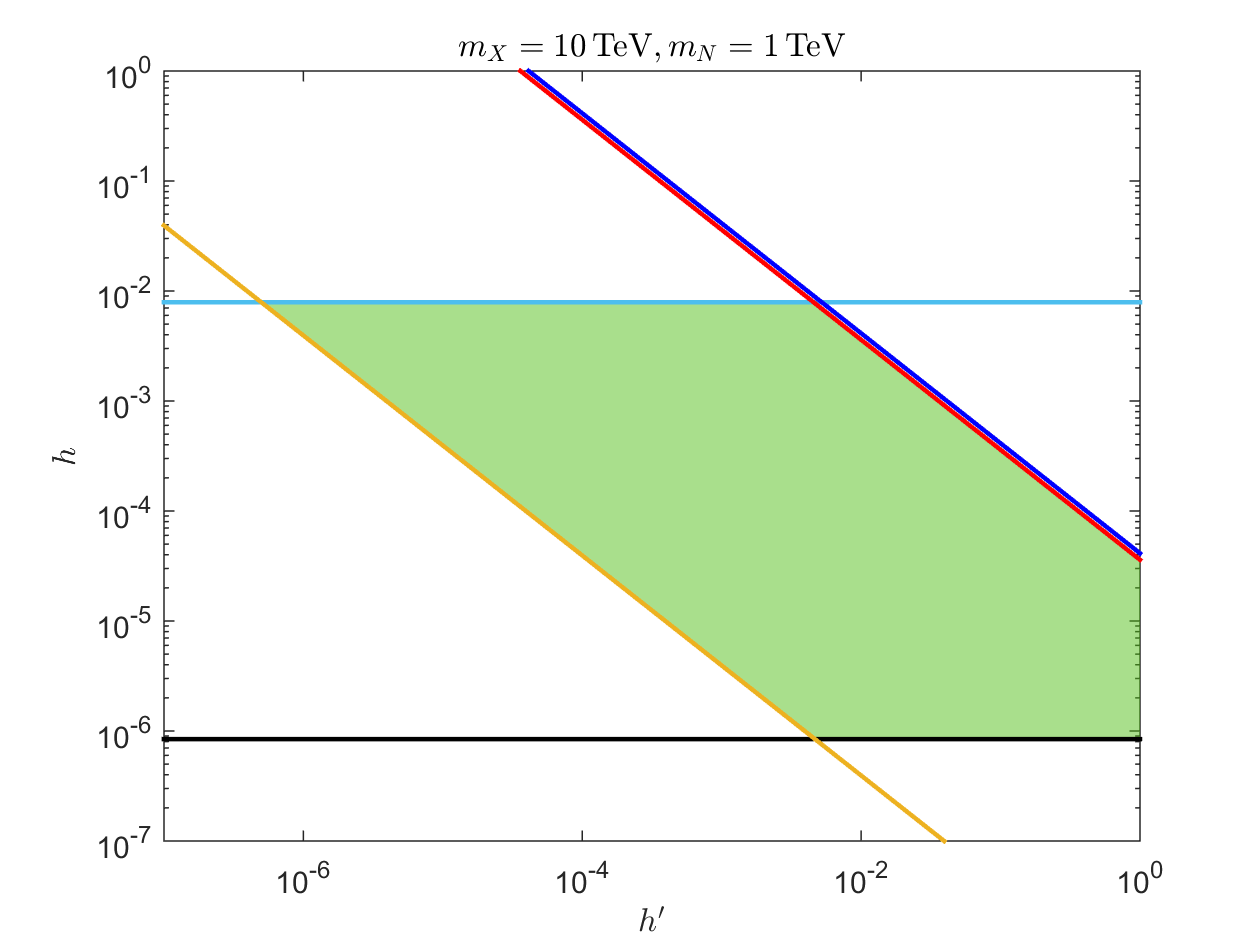}
    \caption{\normalsize Allowed region in the \(h-h'\) plane (shaded green) after imposing the conditions in Eqs.~(\ref{equi},\ref{selfann},\ref{ann},\ref{dom},\ref{onset}). Panels correspond to different values of $m_X$ and $m_N$, while lines correspond to the limiting values of each condition, as labeled in the top-left panel.}
    \label{fig:h_hprime}
\end{figure}

We now impose the conditions given in Eqs.~(\ref{equi},\ref{selfann},\ref{ann},\ref{dom},\ref{onset}) by using the expressions in Eqs.~(\ref{Xdec}--\ref{Ndec}). 
As a proof of concept, we set $C_1 = C_2 =1$ to demonstrate that all of these conditions can be simultaneously satisfied. 
In Fig.~\ref{fig:h_hprime}, we show the allowed region in the $h-h^{\prime}$ plane for various values of $m_X$ and $m_N$ 
that are in the ballpark for testability of the scenario at colliders\footnote{In passing, we note that in the entire allowed regions of the right panels of Fig.~\ref{fig:h_hprime},
and in large parts of the regions in the left panels, the inequality \(\vert h h^{\prime} \vert < 10^{-4}\), mentioned for the explicit model in the previous section, is satisfied.}. 
We see that the conditions in Eqs.~(\ref{ann},\ref{dom}), corresponding to the red and blue lines marked 2 and 1 respectively, essentially lie on top of each other. This can be understood upon substitution of Eqs.~(\ref{Hdom2},\ref{Nann},\ref{Ndec}) as these conditions
have exactly the same dependence on $h$, $h^{\prime}$, $m_X$, and $m_N$ up to an overall numerical factor. 

\begin{figure}[ht!]
    \centering
    \includegraphics[width=0.49\textwidth, trim = .6cm 0cm 1cm .35cm, clip = true]{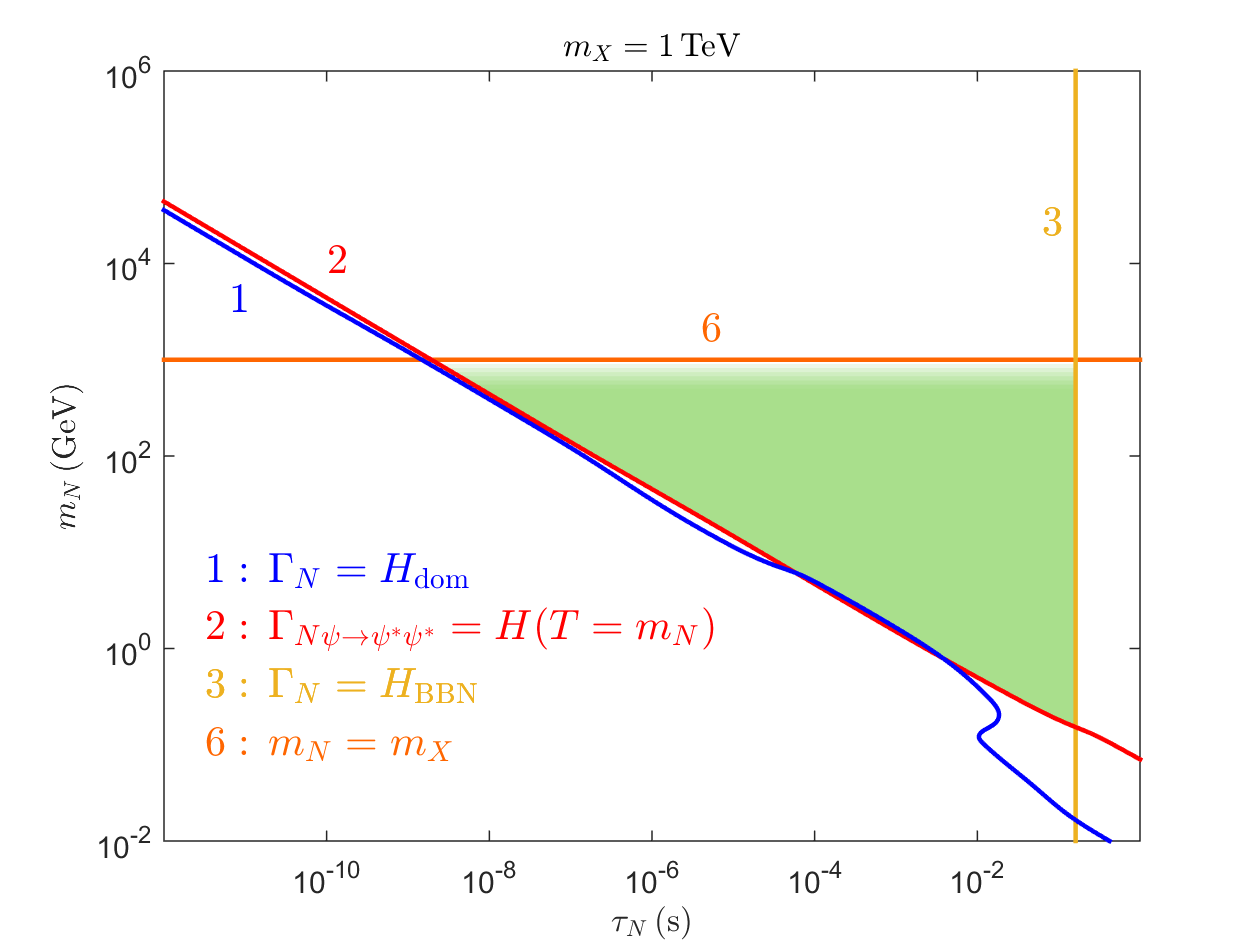}
    \includegraphics[width=0.49\textwidth, trim = .6cm 0cm 1cm .35cm, clip = true]{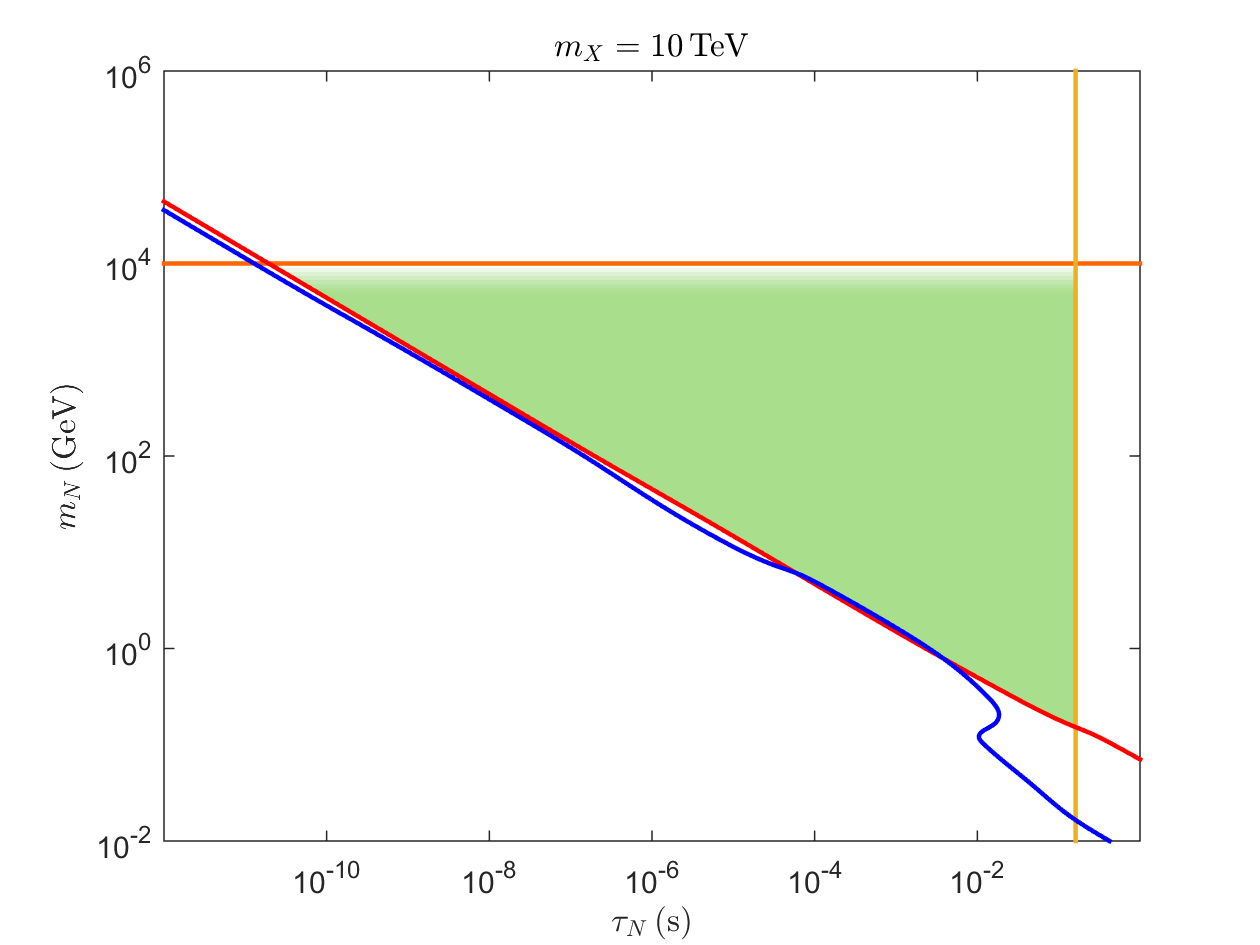}
    \caption{\normalsize Allowed region in the \(m_N-\tau_N\) plane (shaded green) after imposing the conditions in Eqs.~(\ref{ann},\ref{dom},\ref{onset}), as well as \(m_N < m_X\). The various conditions are labeled as in Fig.~\ref{fig:h_hprime}, with the addition of condition 6. The faded portion at the top of the allowed region, just below the \(m_N = m_X\) line,
    emphasizes that \(m_N \ll m_X\) is needed for our scenario. 
    }
    \label{fig:mN_tauN}
\end{figure}

In Fig.~\ref{fig:mN_tauN}, we show the allowed region of the $m_N-\tau_N$ plane for the values of $m_X$ used in Fig.~\ref{fig:h_hprime}, with the same coloring and labeling of the various conditions. We note that the condition in Eq.~(\ref{equi}) has no dependence on \(m_N\) or \(\tau_N\), and when Eq.~(\ref{equi}) is satisfied by even a few orders of magnitude, the condition in Eq.~(\ref{selfann}) is comfortably satisfied as well, as seen in Fig.~\ref{fig:h_hprime}. Therefore we do not show either of these conditions in Fig.~\ref{fig:mN_tauN} (instead, we discuss them in terms of the \(m_N - h\) plane in Appendix C).
The condition in Eq.~(\ref{onset}) is readily translated into an upper bound on $\tau_N \equiv \Gamma^{-1}_N$, and is shown by the yellow vertical line marked 3. The conditions
marked 1 and 2, shown in blue and red, again lie nearly on top of each other for most of the parameter space. However, they now deviate from straight lines 
in the bottom-right of both panels of Fig.~\ref{fig:mN_tauN} due to changes in \(g_*(T)\) (see Appendix B). In particular, as seen in Eq.~(\ref{Hdom2}), the blue curve depends on both \(g_{*N}\) and \(g_{*\rm dom}\) which, along with the sharp drop in \(g_*(T)\) near the QCD phase transition, is responsible for the behavior at low \(m_N\).

One comment is in order at this point. When $m_N$ is very close to $m_X$, the expression in Eq.~(\ref{Xdec}) is no longer a good approximation, and $\Gamma_{X \rightarrow N}$ decreases due to phase-space suppression. Moreover, as shown in Appendix A,
the comoving energy density of $N$ reaches a thermal value and then freezes at a temperature $T \gtrsim m_X/5$ provided that \(m_N \ll m_X\), where the exact temperature depends on the ratio $\Gamma_{X \rightarrow N}/H(T = m_X)$. Together, these considerations imply that as $m_N$ approaches $m_X$, the final abundance of $N$ and hence the onset of the EMD epoch will change as compared to the $m_N \ll m_X$ limit. 
For this reason, the portion of the allowed region in Fig.~\ref{fig:mN_tauN} (as well as in subsequent figures)
that is very close to the $m_N = m_X$ line may be considered ``less viable".



\section{Connection to Experiment}

In this section, we discuss how the scenario may be tested experimentally. Specifically, we explore possible correlations between the allowed regions of the $m_N-\tau_N$ plane and $\langle \sigma_{\rm ann} v \rangle_{\rm f}$ in light of DM indirect detection searches. We also point out the prospects for detecting $N$ via proposed searches for long-lived particles at the LHC.

\subsection{Implications for Dark Matter}

If DM freeze-out occurs after the end of the EMD epoch driven by $N$, then $T_{\rm f} < T_{\rm dec}$ and the standard picture of thermal DM will hold. Otherwise, EMD will affect the DM abundance. If freeze-out happens before the onset of EMD, $T_{\rm f} > T_{\rm dom}$, the relic abundance will be set during RD prior to EMD, and entropy release by $N$ decay will subsequently dilute it. If freeze-out occurs during EMD, $T_{\rm dec} < T_{\rm f} < T_{\rm dom}$, thermal production of DM will also be altered~\cite{GKR,E1}. Moreover, DM particles can be directly produced in the decay of the matter-like component that drives EMD~\cite{KMY,MR,GG,ADS}.

In our scenario, the EMD epoch starts quite late and can end just before the onset of BBN. Eq.~(\ref{Hdom2}) results in $T_{\rm dom} \sim m_N/50$, where we have used $g_* \sim {\cal O}(100)$, whereas $T_{\rm dec} \gtrsim 3$ MeV. On the other hand, $T_{\rm f} \sim m_{\rm DM}/20$. Therefore, unless DM is very light, its relic abundance will be affected by EMD. Below, we discuss the cases with small and large $\langle \sigma_{\rm ann} v \rangle_{\rm f}$ separately.

\vspace{-.2cm}
\subsubsection{$\langle \sigma_{\rm ann} v \rangle_{\rm f} < 3 \times 10^{-26}{\rm cm}^3{\rm s}^{-1}$} 

In a standard thermal history, this leads to overproduction of DM. However, an epoch of EMD can regulate the relic abundance. Let us consider the case where $m_N < m_{\rm DM}$. As mentioned above, this implies that $T_{\rm dom} < T_{\rm f}$, and hence freeze-out occurs in the RD phase preceding the EMD era. Since $N$ is lighter than the DM particle, direct production of DM from $N$ decay is kinematically forbidden. Therefore, the only role of EMD is to dilute the overabundance of DM. The final DM abundance, normalized by the entropy density $s$, then follows:
\begin{equation}
\left({n_{\rm DM} \over s}\right) = d^{-1} \times {3 \times 10^{-26} ~ {\rm cm}^3 ~ {\rm s}^{-1} \over \langle \sigma_{\rm ann} v \rangle_{\rm f}} \times \left({n_{\rm DM} \over s}\right)_{\rm obs}, 
\end{equation}
where the dilution factor $d$ is given by the expression in Eq.~(\ref{d3}), and the observed DM relic abundance is:
\begin{equation} \label{relic1}
\left({n_{\rm DM} \over s}\right)_{\rm obs} \simeq 4.2 \times 10^{-10} \left({1 ~ {\rm GeV} \over m_{\rm DM}}\right).
\end{equation}
Thus, to obtain the correct relic abundance, we need to have:
\begin{equation} \label{cond1}
d = {3 \times 10^{-26} ~ {\rm cm}^3 ~ {\rm s}^{-1} \over \langle \sigma_{\rm ann} v \rangle_{\rm f}}.    
\end{equation}

\begin{figure}[ht!]
    \centering
    \includegraphics[width=0.49\textwidth, trim = .6cm 0cm 1cm 0cm, clip = true]{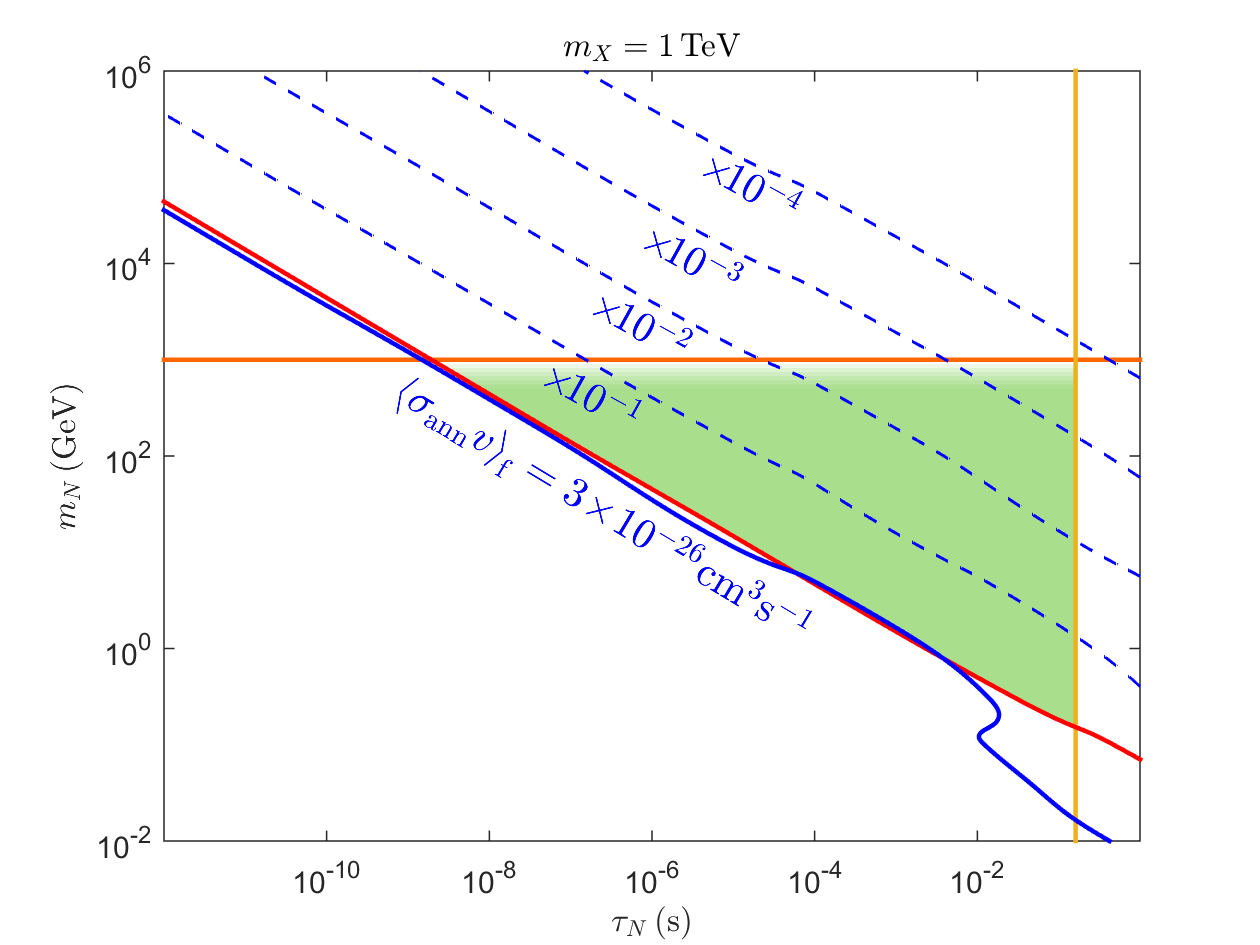}
    \includegraphics[width=0.49\textwidth, trim = .6cm 0cm 1cm 0cm, clip = true]{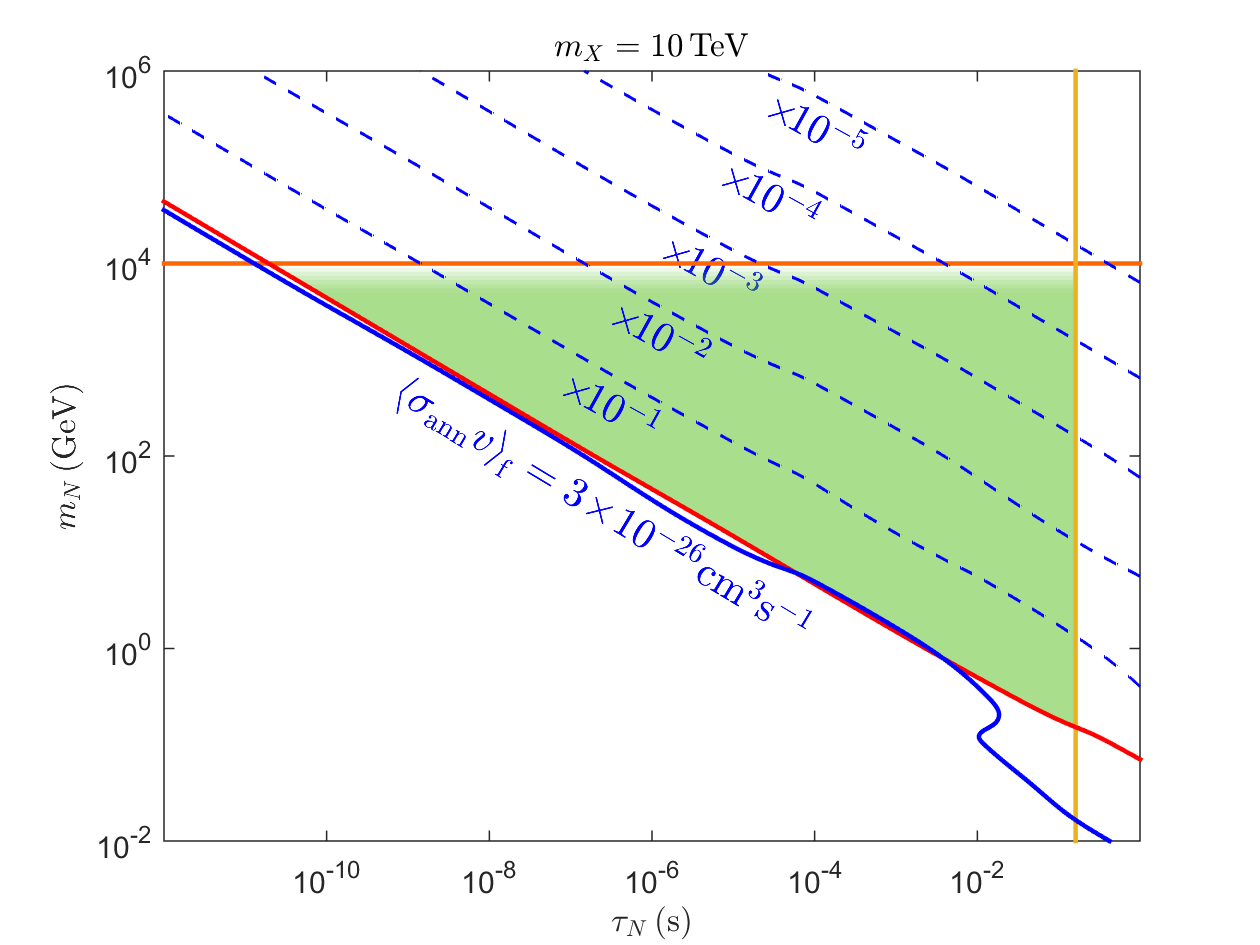}
    \caption{\normalsize Contours of \(\left<\sigma_{\rm ann} v\right>_{\rm f}\) that reproduce the observed DM abundance, shown as blue dashed curves, in the \(m_N - \tau_N\) plane for the case when \(\left<\sigma_{\rm ann} v\right>_{\rm f} < 3\times 10^{-26}{\rm cm}^3{\rm s}^{-1}\). The shaded region and its boundaries are those of Fig.~\ref{fig:mN_tauN}. When \(\left<\sigma_{\rm ann} v\right>_{\rm f} = 3 \times 10^{-26}\) cm$^3$ s$^{-1}$
    , the contour coincides with the solid blue curve of Fig.~\ref{fig:mN_tauN}. This essentially corresponds to the absence of EMD as \(\Gamma_N = H_{\rm dom}\) in this case, thus recovering the standard freeze-out scenario. 
    As before, the temperature dependence of \(g_*(T)\) is responsible for deviations from straight lines. }
    \label{fig:DM_small}
\end{figure}

From Eqs.~(\ref{Hdom2},\ref{d3}), we see that $d$ is basically determined once we know $m_N$ and $\Gamma_N$. Using this, in Fig.~\ref{fig:DM_small} we plot contours of $\langle \sigma_{\rm ann} v \rangle_{\rm f}$ in the $m_N - \tau_N$ plane that yield the observed DM abundance for the same values of $m_X$ as in Fig.~\ref{fig:mN_tauN}. The contours show \(\langle \sigma_{\rm ann} v \rangle_{\rm f}\) decreasing by factors of 10, relative to the nominal value of $3 \times 10^{-26}$ cm$^3$ s$^{-1}$, corresponding to successively longer periods of EMD with more and more dilution. The region of interest is where the dashed contours overlap with the green allowed region of our scenario. 
We see that the scenario can give the desired relic abundance for $\langle \sigma_{\rm ann} v \rangle_{\rm f}$ as small as $\sim 10^{-30}\,{\rm cm}^3{\rm s}^{-1}$, which is well below the current Fermi-LAT bounds from observations of dwarf spheroidal galaxies~\cite{fermi1} and newly discovered Milky Way satellites~\cite{fermi2}. 

\vspace{-.3cm}
\subsubsection{$\langle \sigma_{\rm ann} v \rangle_{\rm f} > 3 \times 10^{-26}{\rm cm}^3{\rm s}^{-1}$} 

Freeze-out in a standard thermal history, as well as in an EMD scenario, leads to an underabundance of DM in this case. However, direct production of DM from $N$ decay combined with a large value of $\langle \sigma_{\rm ann} v \rangle_{\rm f}$ can yield the correct relic abundance. If $N$ decay results in a DM density that is higher than the observed value, and $\langle \sigma_{\rm ann} v \rangle_{\rm f}$ is sufficiently large, residual annihilation at the end of the EMD epoch (when $H \simeq \Gamma_N$) gives~\cite{KMY,MR,Scott1,DLS,Scott3}:
%
%
%
\begin{equation} \label{relic2}
\left({n_{\rm DM} \over s}\right) = {T_{\rm f} \over T_{\rm dec}} \times {3 \times 10^{-26} ~ {\rm cm}^3 ~ {\rm s}^{-1} \over \langle \sigma_{\rm ann} v \rangle_{\rm f}} \times \left({n_{\rm DM} \over s}\right)_{\rm obs}.
\end{equation}
This matches the observed DM abundance provided that:
\begin{equation} \label {cond2}
T_{\rm dec} = {3 \times 10^{-26} ~ {\rm cm}^3 ~ {\rm s}^{-1} \over \langle \sigma_{\rm ann} v \rangle_{\rm f}} \times T_{\rm f}.
\end{equation}
We note that $\langle \sigma_{\rm ann} v \rangle_{\rm f} > 3 \times 10^{-26}$ cm$^3$ s$^{-1}$ is allowed within the $20 ~ {\rm GeV} \lesssim m_{\rm DM} \lesssim 100$ TeV mass range. For DM masses up to a few TeV, the upper limit on $\langle \sigma_{\rm ann} v \rangle_{\rm f}$ is set by a recent analysis~\cite{Beacom} of Fermi results (unless there is $P$-wave annihilation or coannihilation), while for larger values of DM mass, it is set by the well-known unitarity bound~\cite{GK}. The largest allowed value is $\langle \sigma_{\rm ann} v \rangle_{\rm f} \simeq 2 \times 10^{-23}$ cm$^3$ s$^{-1}$, which happens at the intersection of these constraints.

\begin{figure}[hp]
    \centering
    \includegraphics[width=0.49\textwidth, trim = .6cm .1cm 1cm .35cm, clip = true]{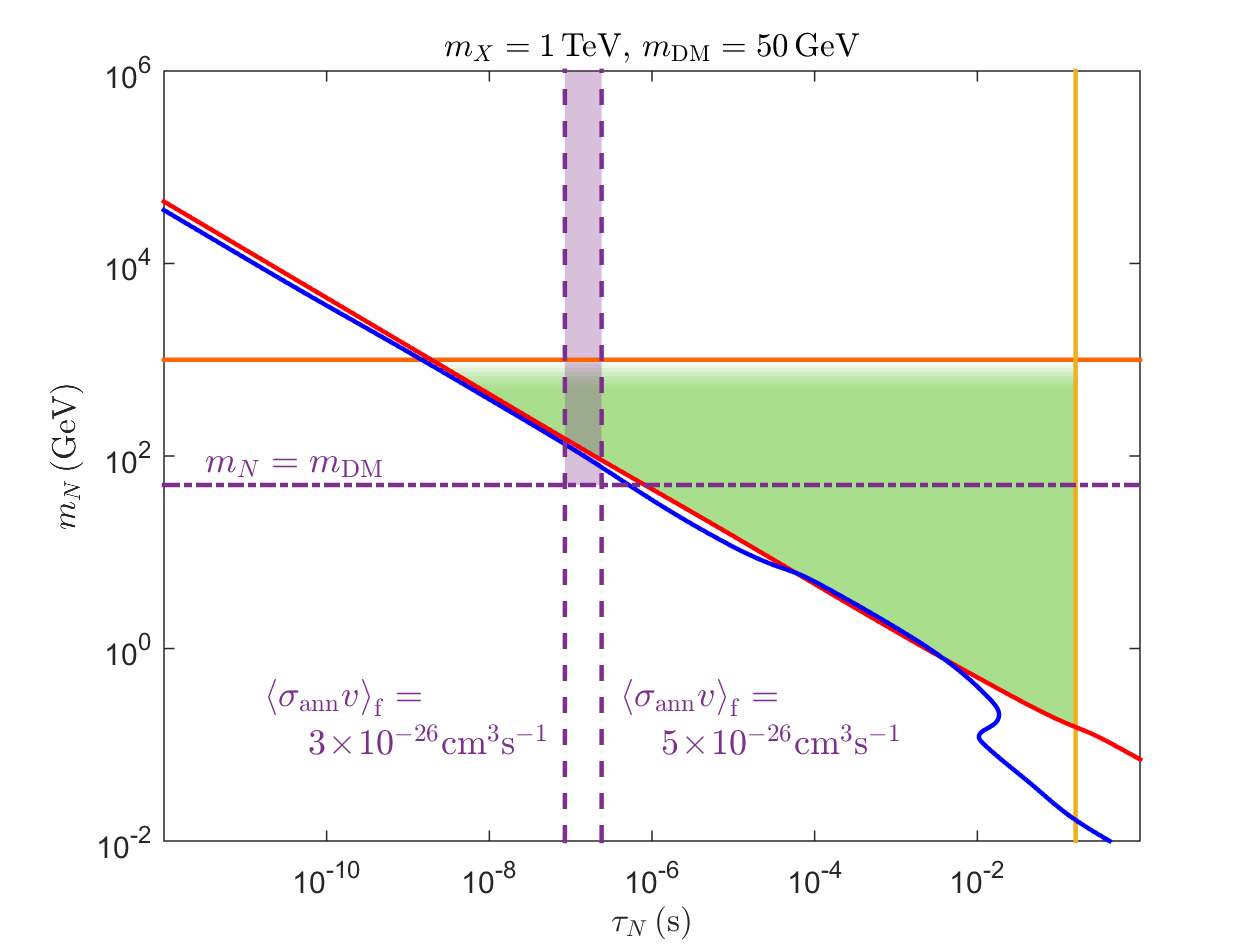}
    \includegraphics[width=0.49\textwidth, trim = .6cm .1cm 1cm .35cm, clip = true]{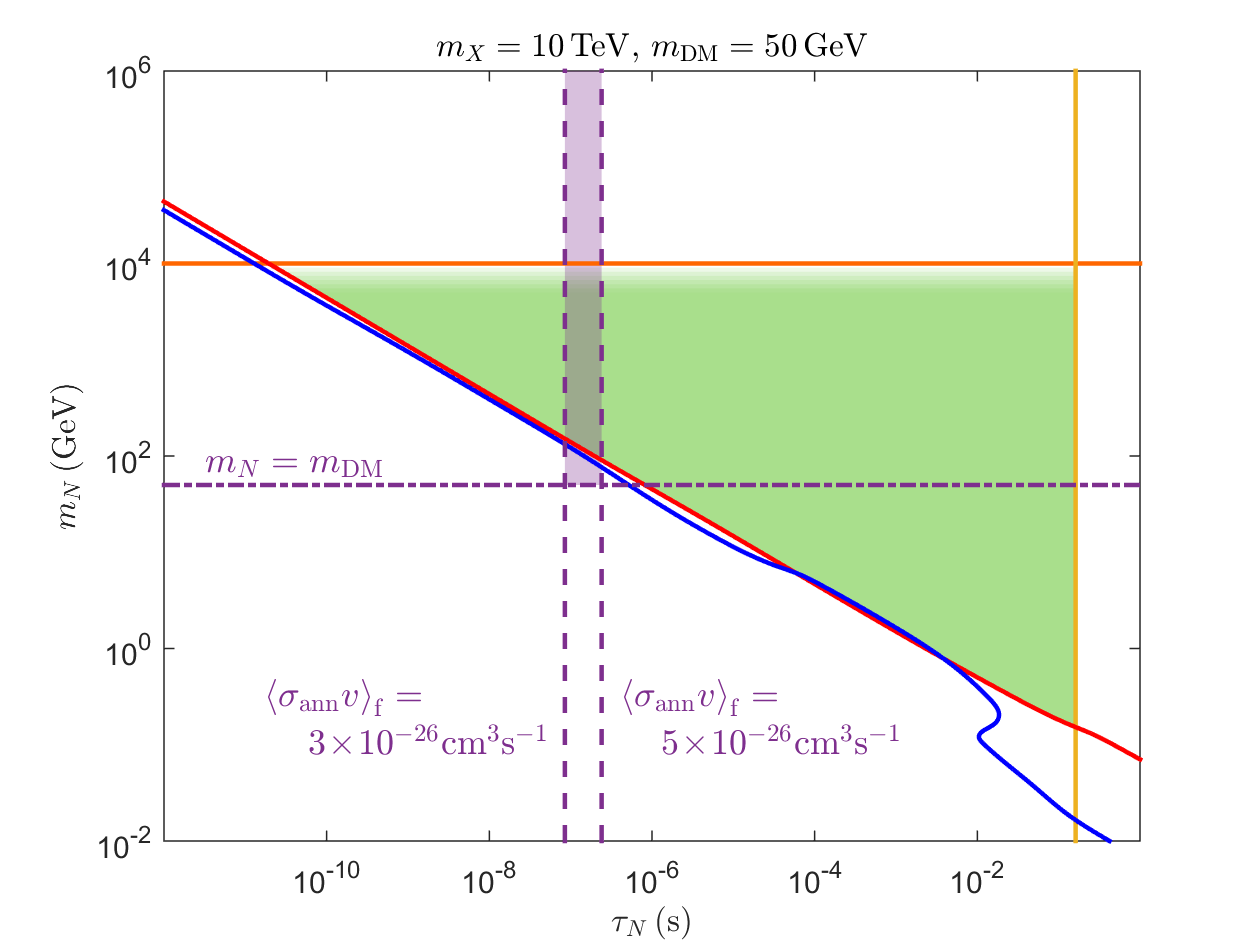}
    \includegraphics[width=0.49\textwidth, trim = .6cm .1cm 1cm .35cm, clip = true]{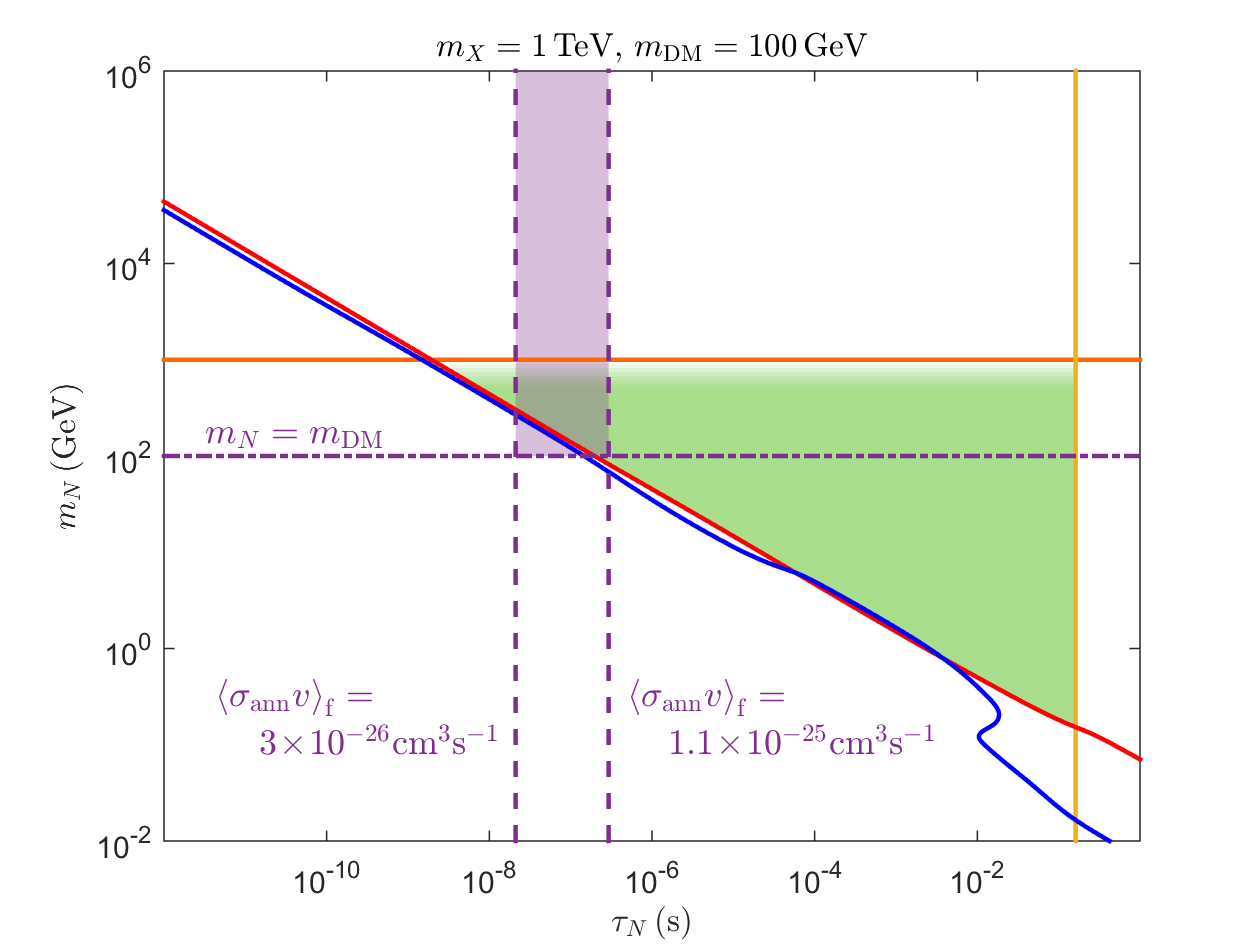}
    \includegraphics[width=0.49\textwidth, trim = .6cm .1cm 1cm .35cm, clip = true]{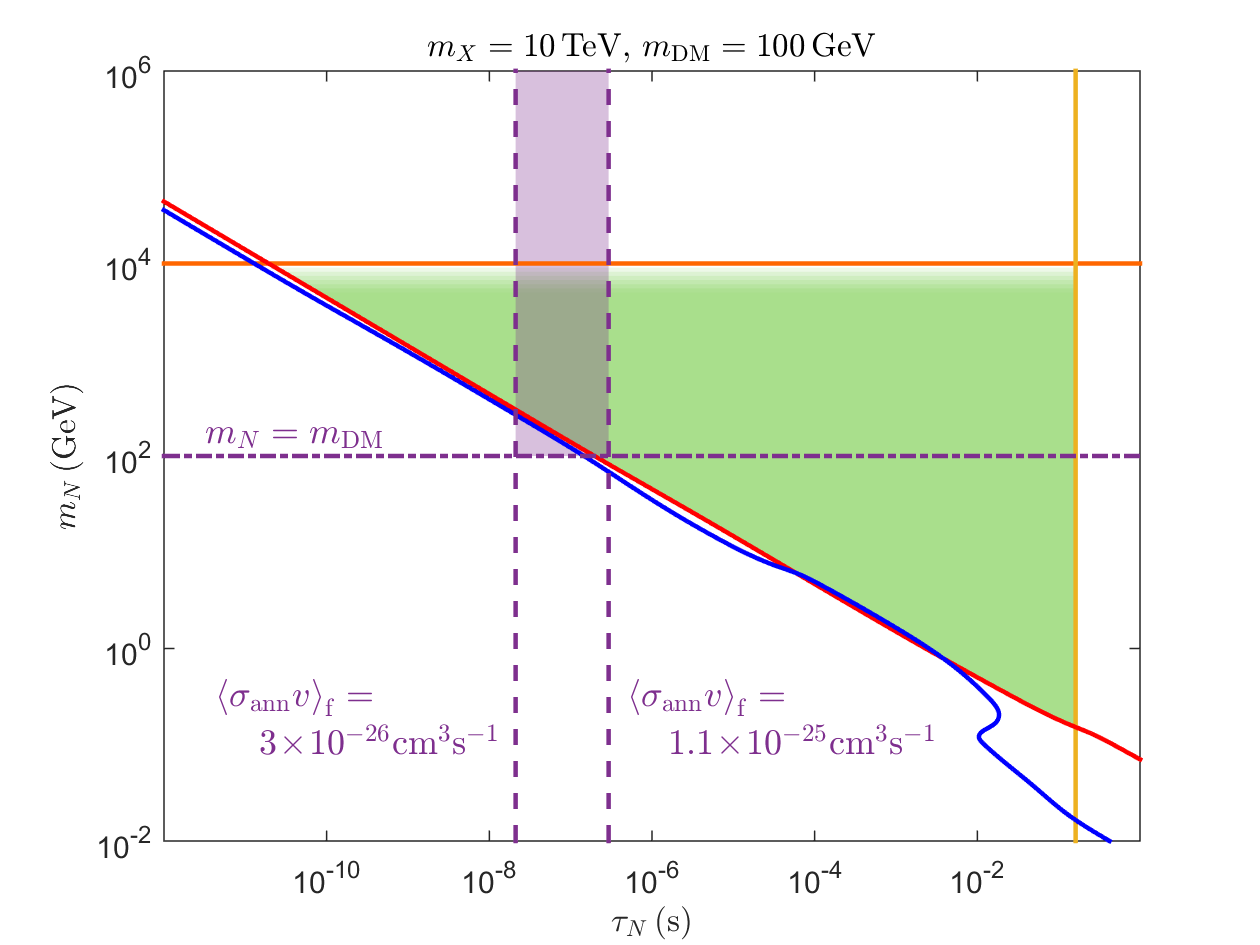}
    \includegraphics[width=0.49\textwidth, trim = .6cm .1cm 1cm .35cm, clip = true]{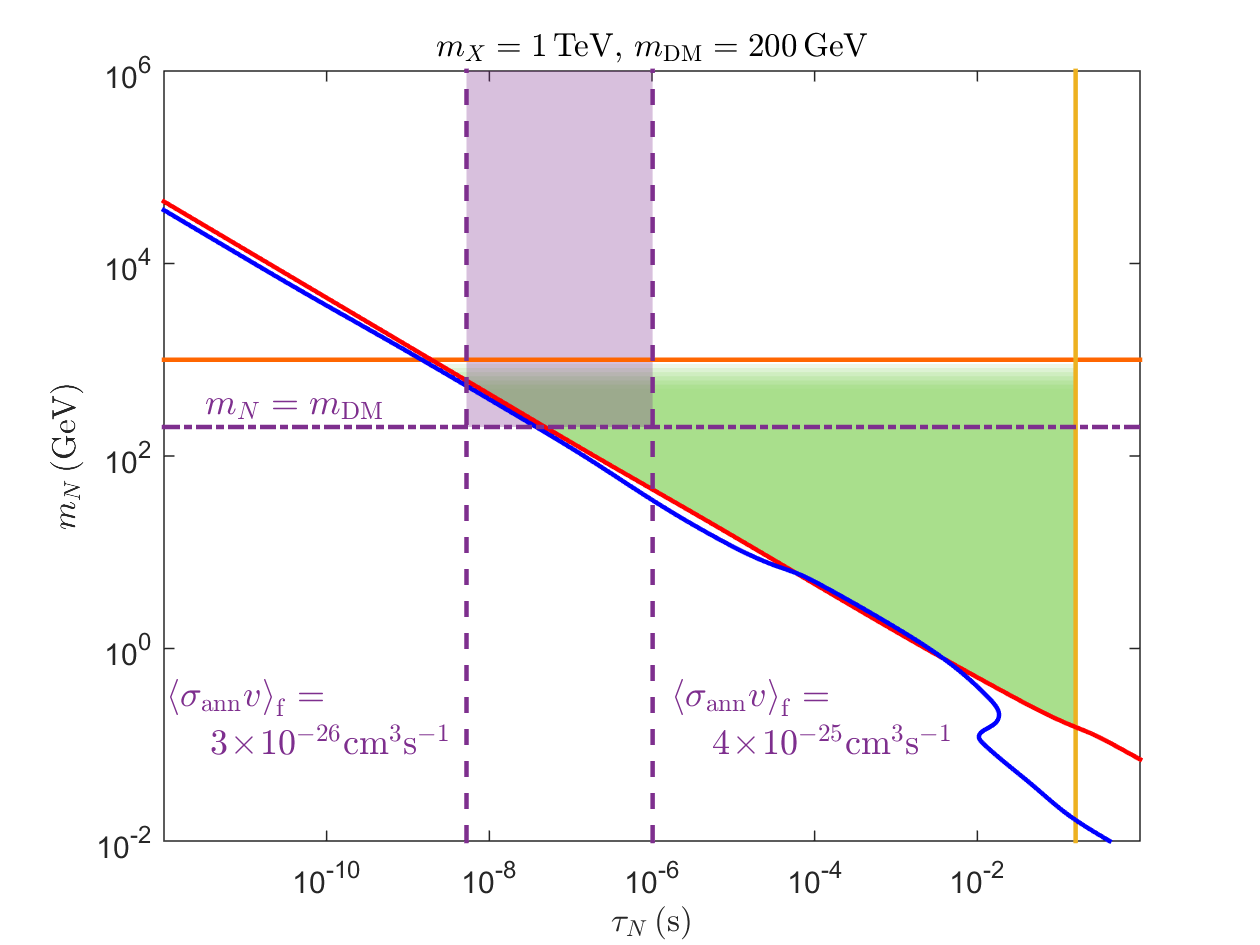}
    \includegraphics[width=0.49\textwidth, trim = .6cm .1cm 1cm .35cm, clip = true]{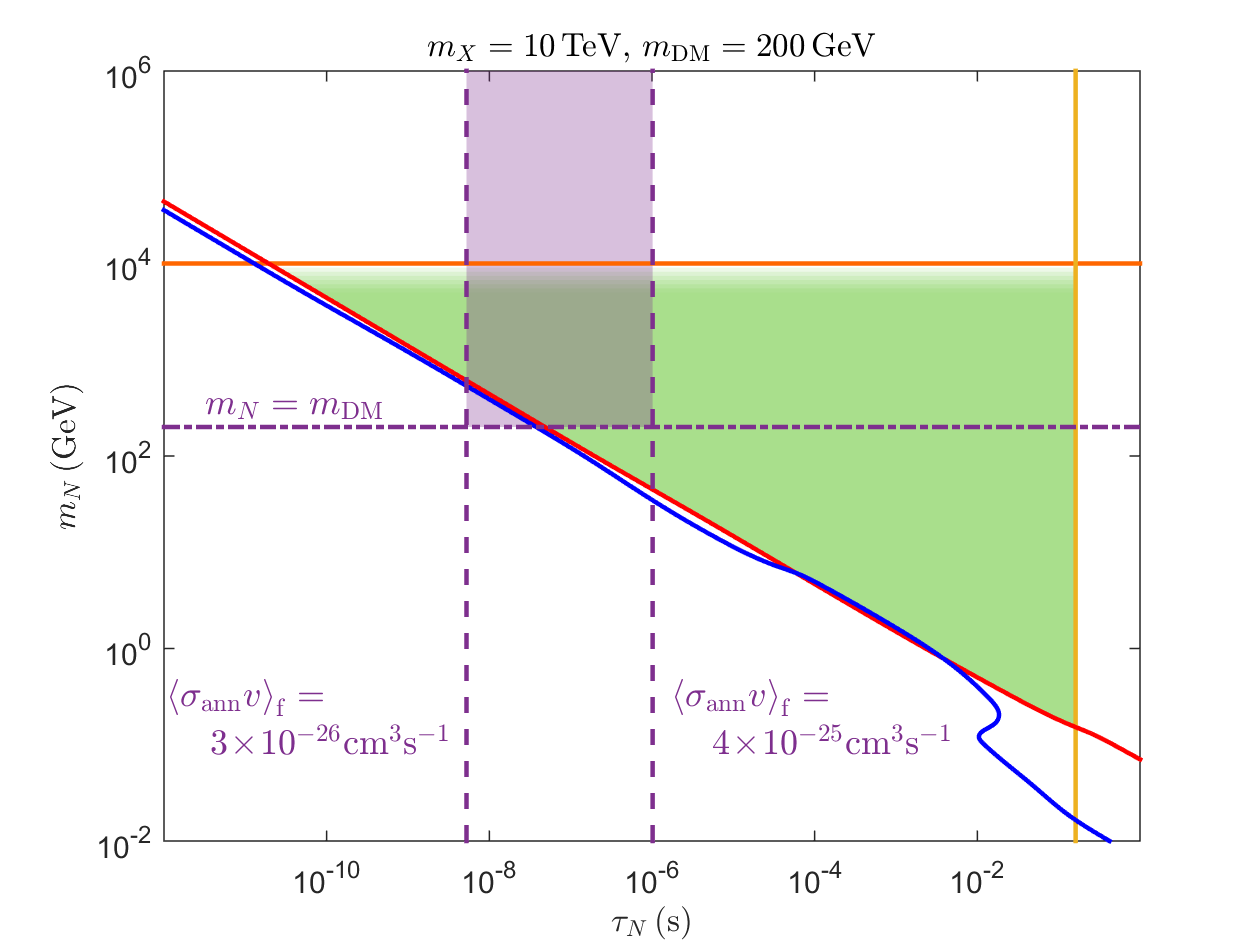}
    \vspace{-0.3cm}
    \caption{\normalsize Regions that yield the correct DM abundance in the \(m_N - \tau_N\) plane for the case when \(\left<\sigma_{\rm ann} v\right>_{\rm f} > 3\times 10^{-26}{\rm cm}^3{\rm s}^{-1}\). Each panel corresponds to fixed \(m_X\) and \(m_{\rm DM}\), and shows the allowed region of Fig.~\ref{fig:mN_tauN}. In each panel, the purple vertical shaded region is restricted from below by \(m_N > m_{\rm DM}\), from the left by \(\left<\sigma_{\rm ann} v\right>_{\rm f} = 3\times 10^{-26}{\rm cm}^3{\rm s}^{-1}\), and from the right by the maximum value of \(\left<\sigma_{\rm ann} v\right>_{\rm f}\) allowed by the bounds in \cite{Beacom} for the given value of \(m_{\rm DM}\), as shown. 
    The region of interest is where the two shaded areas overlap.
    }
    \label{fig:DM_large}
\end{figure}

Fig.~\ref{fig:DM_large} depicts regions in the $m_N-\tau_N$ plane that yield the correct relic abundance for a given $m_{\rm DM}$. We use the allowed range for $\langle \sigma_{\rm ann} v \rangle_{\rm f}$ from combined experimental and unitarity bounds in~\cite{Beacom} for a given $m_{\rm DM}$, and translate it into a range for $T_{\rm dec}$ using Eq.~(\ref{cond2}) with $T_{\rm f} \sim m_{\rm DM}/20$. $T_{\rm dec}$ is readily related to $\Gamma_N$ which then restricts $\tau_N$ to be within a vertical band. By imposing $m_N > m_{\rm DM}$, so that $N$ decay to DM is kinematically allowed, we find the allowed regions shown in purple. Different panels in Fig.~\ref{fig:DM_large} correspond to different values of $m_X$ and $m_{\rm DM}$, while different values of $\langle \sigma_{\rm ann} v \rangle_{\rm f}$ within the allowed range scan horizontally through the vertical bands. In each panel, the region of interest is where the vertical band overlaps with the allowed region of our scenario. The only dependence on \(m_X\) is in the orange horizontal line corresponding to \(m_N = m_X\). As \(m_X\) increases, this line moves upward resulting in more overlap between the two shaded regions.
%

\subsubsection{Other Considerations}

The EMD epoch driven by $N$ can also regulate the DM relic abundance in cases that DM production is not associated with processes in the thermal bath. An important example is axion DM in the form of coherent oscillations of the QCD axion arising due to an initial displacement from its minimum. EMD dilutes axionic DM if $T_{\rm dec} < \Lambda_{\rm QCD} \simeq 200$ MeV thereby avoiding any overproduction~\cite{FPT}. The maximum dilution is obtained for $T_{\rm dec} \simeq T_{\rm BBN} \simeq 3$ MeV, which allows an axion decay constant of order $f_a \simeq  10^{14}$ GeV without fine tuning of the initial misalignment angle. In supersymmetric extensions of the SM, this can be achieved with dilution from the decay of the axino and/or saxion~\cite{Howie1,Howie2}, however, our scenario works just as well in a nonsupersymmetric set up.

Our scenario can be particularly helpful with DM at the high end of the mass spectrum, i.e., $m_{\rm DM} \gg {\cal O}({\rm TeV})$. In this case, according to Eq.~(\ref{relic1}), obtaining the observed relic abundance requires a very small value of $n_{\rm DM}/s$, which could pose a challenge to any production mechanism of superheavy DM. However, entropy release by $N$ decay can dilute the existing DM density by a few orders of magnitude thereby enlarging the parameter space of any model with extremely heavy DM candidates (for example, see~\cite{ABCO}).

Moreover, the EMD epoch driven by $N$ dilutes any relic abundance produced at prior stages of the cosmological history. A notable example is the observed matter-antimatter asymmetry of the Universe. Some mechanisms, notably Affleck-Dine baryogenesis~\cite{Dine1,Dine2}, can generate an excessively large baryon asymmetry. Entropy released in $N$ decay can help regulate the baryon asymmetry in such cases. The same applies to dangerous relics such as unstable gravitinos of weak-scale mass whose abundance is constrained by the success of BBN~\cite{Gravitino1,Gravitino2}.

\subsection{Testable Collider Signals}

Due to its long lifetime, required for the success of our scenario, $N$ is an example of a long-lived particle (LLP). This provides us with an opportunity to directly probe a sub-TeV $N$ via the proposed LLP searches at the LHC. As we have seen, the allowed region in the $m_N-\tau_N$ plane covers a wide range of $N$ lifetime. The rest-frame lifetime $\tau_N$ of $N$ particles produced from $X$ decay in colliders is related to the decay length $l_N$ according to $l_N = {\bar b} c \tau_N$. 
Here, ${\bar b}$ is the average boost factor of $N$, which we take to be ${\bar b} \sim m_X/2 m_N$. Decay lengths above $10^{-2}$ cm are within the reach of the LHC. For $10^{-2} ~ {\rm cm} < l_N < 10^2$ cm, $N$ production and decay gives rise to displaced vertices at the LHC, while the case with $10^2 ~ {\rm cm} < l_N < 10^4$ cm leads to displaced jet/lepton signals. 

However, neutral LLPs (like our $N$) with decay lengths above 100 m are particularly difficult to probe because of the limited sensitivity of the LHC main detectors.
The recently proposed MATHUSLA (MAssive Timing Hodoscope for Ultra Stable neutraL pArticles) detector concept~\cite{MAT1} is a minimally instrumented, large-volume surface detector located near ATLAS or CMS. It would search for neutral LLPs produced in the high luminosity LHC (HL-LHC) collisions, extending the lifetime range by a few orders of magnitude compared to the main detectors. It could discover LLPs with decay lengths up to $3 \times 10^7$ m, which correspond to lifetimes close to the age of the Universe at the onset of BBN~\cite{MAT2}.        

The case when $l_N > 100$ m is particularly interesting as it may be probed by ATLAS and CMS as well as MATHUSLA. 
To be specific, let us focus on the model in Eq.~(\ref{lagran}) with $m_X \sim 1-10$ TeV and $m_N \sim 100 ~ {\rm GeV}-1$ TeV as a canonical example. $X$ can be produced at the LHC due to its couplings to the quarks in this model, and its decay yields dijet and monojet signals (the latter accompanied by $N$). These signals have been studied in detail in~\cite{LHC1,ADD}, and in a recent analysis by CMS~\cite{CMS}, when $N$ is absolutely stable and $m_N \approx 1$ GeV (hence light DM). One may generalize these studies to $m_N \sim 100 ~ {\rm GeV}-1$ TeV in a rather straightforward manner. 

Interestingly, our benchmark point also overlaps with MATHUSLA’s most important physics target, namely hadronically decaying LLPs with masses in the ${\cal O}(10 ~ {\rm GeV}) - {\cal O}(100 ~ {\rm GeV})$ range~\cite{MAT3}. The collaboration has studied the discovery potential of such LLPs produced from exotic Higgs decays~\cite{MAT2}, or through mixing of the (scalar) LLP with Higgs~\cite{MAT3}. In our model, as mentioned, the main production channel for $N$ is decay of $X$ particles. This should be implemented properly to calculate the cross-section for $N$ production. Thus, combining MATHUSLA with the main detectors will allow us to probe the $m_N-\tau_N$ plane at the HL-LHC. 

\begin{figure}[ht!]
    \centering
    \includegraphics[width=0.49\textwidth, trim = .6cm 0cm 1cm .35cm, clip = true]{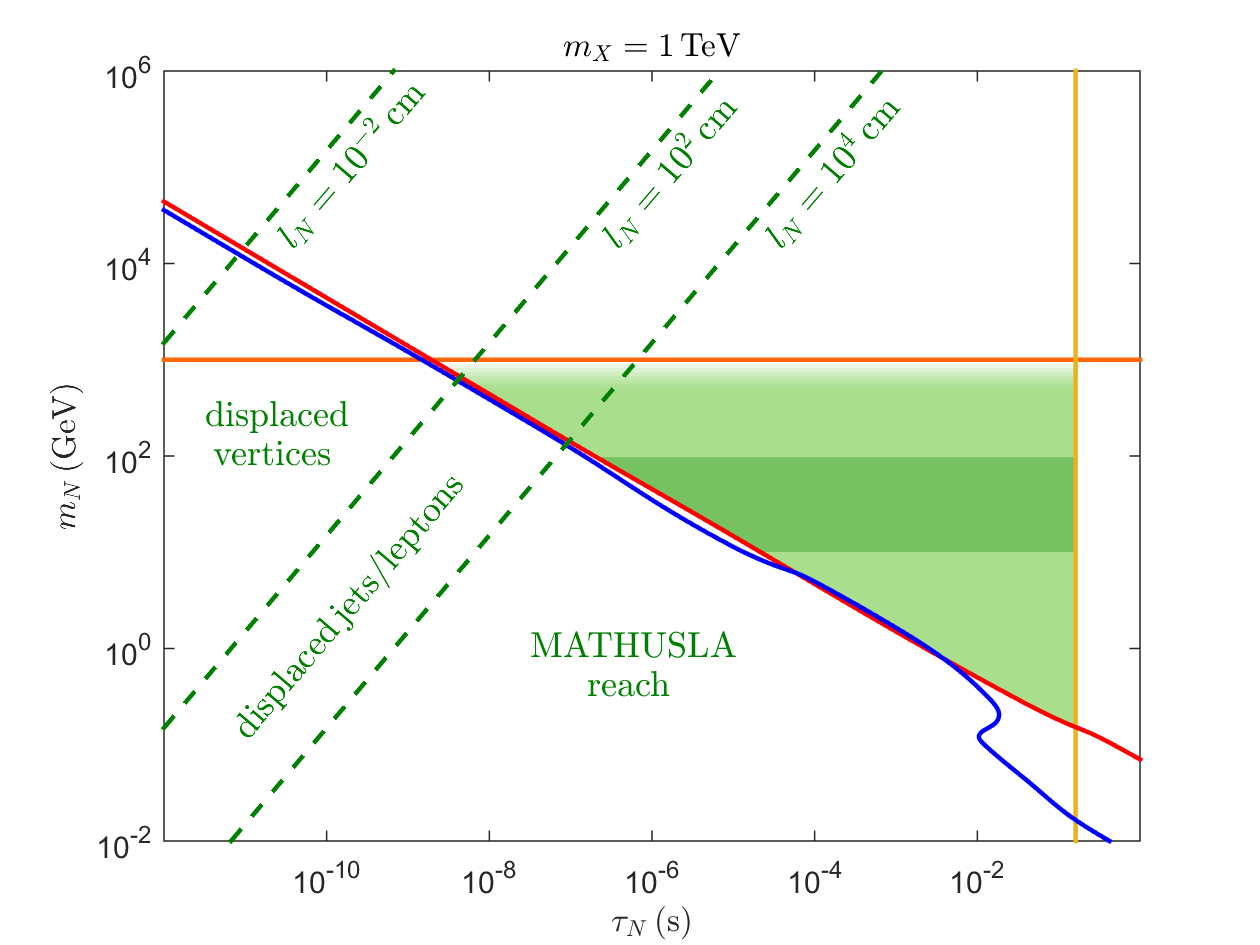}
    \includegraphics[width=0.49\textwidth, trim = .6cm 0cm 1cm .35cm, clip = true]{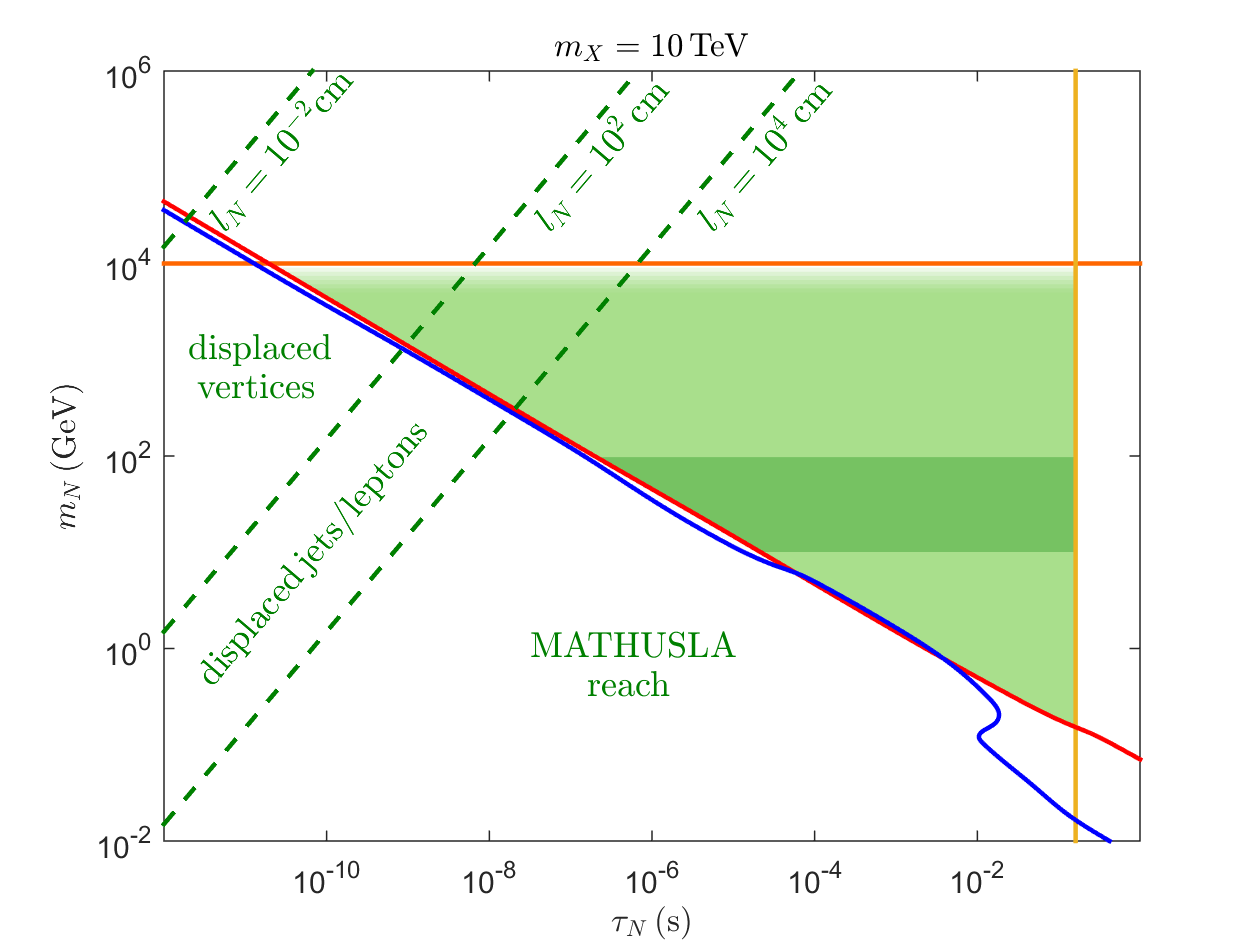}
    \caption{\normalsize The allowed region in the \(m_N - \tau_N\) plane, separated by collider signatures based on the decay length of \(N\). The diagonal band between \(10^{-2}\) cm \(< l_N < 10^2\) cm 
    leads to displaced vertices at the LHC. The next band, with \(10^2\) cm \(< l_N < 10^4\) cm, leads to displaced jet/lepton signals. Finally, the right-most region, with \(l_N > 10^4\) cm, would be probed by MATHUSLA. The darker shaded band within the allowed region of our scenario corresponds to \(m_N = 10\) GeV\(- 100\) GeV, which is the most important MATHUSLA target for hadronically decaying LLPs.}
    \label{fig:collider}
\end{figure}

Fig.~\ref{fig:collider} shows the allowed region of our scenario in the \(m_N - \tau_N\) plane with regimes of the decay length and corresponding prospects for detection at the LHC. Though part of our allowed region extends to \(l_N < 10^2\) cm, most of it lies at longer lifetimes with \(l_N > 10^4\) cm. We also see that the region relevant for MATHUSLA has some overlap with the regions that yield the correct DM abundance for $\langle \sigma_{\rm ann} v \rangle_{\rm f} > 3 \times 10^{-26}$ cm$^3$ s$^{-1}$ in Fig.~\ref{fig:DM_large}. The situation is even better, see Fig.~\ref{fig:DM_small}, in the case that $\langle \sigma_{\rm ann} v \rangle_{\rm f} < 3 \times 10^{-26}$ cm$^3$ s$^{-1}$. Note that we have expressed the average boost factor of $N$ particles as ${\bar b} \sim m_X/2m_N$, which is strictly correct when $X$ decays at rest. However, $X$ particles produced at the LHC are boosted themselves, leading to a larger ${\bar b}$ for $N$. In this case, the lines corresponding to different values of $l_N$ would move up and to the left, making the region within the reach of MATHUSLA even larger.  

Potential collider measurement of $m_N$ and $\tau_N$ has very interesting cosmological implications. $\tau_N$ (equivalently $\Gamma_N$) is a direct measure of the temperature $T_{\rm dec}$ when $N$ decay reheats the Universe. This leads to the tantalizing possibility of determining the highest temperature of the Universe in the last phase of RD, which is relevant for BBN, via particle physics experiments. By knowing $m_N$, we can also find the onset of the EMD period that is driven by $N$ through Eq.~(\ref{Hdom2}). Therefore, in principle, we can directly probe the entire EMD epoch ending just before the onset of BBN with collider experiments. 

\vspace{-0.2cm}
\section{
Conclusion}

In this paper, we presented a nonstandard cosmological history scenario where a visible-sector particle drives a period of EMD. This scenario involves a sub-TeV SM singlet $N$ that acquires a thermal abundance at temperatures well above its mass as a result of the decay/inverse decay of a parent particle $X$ with SM charge(s). Then, being long lived, $N$ dominates the energy density of the Universe as a frozen species and leads to an EMD epoch that can last until the onset of BBN. The scenario works for fermionic and bosonic $N$ equally well. Moreover, since the energy density of $N$ originates from the thermal bath, it automatically evades isocurvature bounds. We discussed an explicit realization of the scenario where $N$ is a Majorana fermion coupled to quarks. 

We outlined the necessary conditions in order for $N$ to reach and maintain an equilibrium energy density, dominate the Universe, and decay before the onset of BBN. We showed that these conditions can be simultaneously satisfied in significant parts of the parameter space. It is possible to obtain the correct DM relic abundance within this allowed parameter space for both cases with $\langle \sigma_{\rm ann} v \rangle_{\rm f} < 3 \times 10^{-26}$ cm$^3$ s$^{-1}$ and $\langle \sigma_{\rm ann} v \rangle_{\rm f} > 3 \times 10^{-26}$ cm$^3$ s$^{-1}$. Moreover, the entropy release from $N$ decay can regulate baryon asymmetry generated at earlier stages of the cosmological history and dilute the abundance of dangerous relics like unstable gravitinos. 

An interesting aspect of this scenario is that the $m_N-\tau_N$ plane may be directly probed by collider experiments. In large parts of the allowed parameter space, $\tau_N$ is in the range for discovery by the MATHUSLA proposal at the HL-LHC. This is particularly the case for a hadronically decaying $N$ with $m_N \sim {\cal O}(100 ~ {\rm GeV})$, which is MATHUSLA's most important physics target. Measuring $\tau_N$ and $m_N$ exhibits a profound interplay between particle physics and cosmology. Determining $\tau_N$ will give us the highest temperature in the RD phase that sets the stage for BBN. Using information about $m_N$, in tandem, we can fully reconstruct the EMD era driven by $N$. 

In summary, a visible sector particle $N$ with sub-TeV mass can give rise to a (final) period of EMD in the early Universe rather naturally. This scenario is robust, and largely independent from the details of the earlier stages of the cosmological history, as long as the Universe is in a RD phase at $T \gg m_N$. A logical next step is
embedding the scenario in realistic extensions of the SM with TeV-scale long-lived fermions or scalars.
Another important direction is to employ exciting rapid developments in LLP searches and perform detailed analysis of the discovery prospect of such models at the energy frontier. 
We leave these investigations for future work.

\section*{Acknowledgements}


JO is supported by the project AstroCeNT: Particle Astrophysics Science and Technology Centre, carried out within the International Research Agendas programme of the Foundation for Polish Science financed by the European Union under the European Regional Development Fund.

\section*{
}

\vspace{-.8cm}
\subsection{Evolution of the Energy Density}

Here, we would like to derive the time evolution of the energy density of $N$ particles produced via the interaction term $h X N \psi$ in Eq.~(\ref{Lnew}). We work in the limit $m_\psi \ll m_N \ll m_X$. For simplicity, we take $m_N = m_\psi = 0$ below. This is a good approximation as long as we are interested in the evolution of $\rho_N$ over time scales where $T \gg m_N$.

Let us start with the equation that governs the occupation number of $N$, denoted by $f_N$:
\begin{equation} \label{EE1}
{df_N({\vec p}_N) \over dt} = \int{{d^3 p_X \over (2 \pi)^3} {d^3 p_\psi \over (2 \pi)^3} {\vert {\cal M} \vert^2 \over 8 E_X E_\psi E_N} (f_X({\vec p}_X) - f_\psi({\vec p}_\psi) f_N({\vec p}_N)) (2 \pi)^4 \delta^{(3)} ({\vec p}_X - {\vec p}_N - {\vec p}_\psi) \delta(E_X - E_N - E_\psi)} .
\end{equation}
Here, $f_X$ and $f_\psi$ are the occupation numbers of $X$ and $\psi$ respectively, and 
${\cal M}$ is the Feynman amplitude for the $X \rightarrow N \psi^*$ process (note that, being a Majorana fermion, $N$ is its own antiparticle). Since $X$ and $\psi$ are in thermal equilibrium, $f_X = f_X^{\rm eq}$ and $f_\psi = f_\psi^{\rm eq}$. We also have $\vert {\cal M} \vert = h m_X$. 

Integration over ${\vec p}_\psi$ yields:
\begin{equation} \label{EE2}
{df_N({\vec p}_N) \over dt} = \int{{d^3 p_X \over (2 \pi)^2} {h^2 m^2_X \over 8 E_X E_N E_\psi} (f_X({\vec p}_X) - f_\psi({\vec p}_\psi) f_N({\vec p}_N)) \delta(E_X - E_N - E_\psi)},
\end{equation}
where:
\begin{equation}
E_X = \sqrt{\vert {\vec p}_X \vert^2 + m^2_X} ~ ~ ~ , ~ ~ ~ E_N = \vert {\vec p}_N \vert ~ ~ ~ , ~ ~ ~ E_\psi = \vert {\vec p}_\psi \vert .
\end{equation}
Conservation of momentum implies that:
\begin{equation} \label{mom1}
E_\psi = \sqrt{\vert {\vec p}_X \vert^2 + \vert {\vec p}_N \vert^2 - 2 {\vec p}_X \cdot {\vec p}_N}.
\end{equation}
Using spherical coordinates, with the $z$ axis chosen in the direction of ${\vec p}_N$, we have:
\begin{equation} \label{mom2}
E_\psi = \sqrt{p^2_X  + p^2_N - 2 p_X p_N {\rm cos} \theta},
\end{equation}
where $p_X \equiv \vert {\vec p}_X \vert$ and $p_N \equiv \vert {\vec p}_N \vert$. Integrating over the solid angle, and noting isotropy about ${\vec p}_N$, we find:
\begin{equation} \label{EE3}
{df_N({\vec p}_N) \over dt} = {m_X \Gamma_{X \rightarrow N} \over p^2_N} (1 -  e^{\beta p_N} f_N({\vec p}_N)) \int{{e^{- \beta E_X} p_X \over E_X } d p_X},
\end{equation}
where $\Gamma_{X \rightarrow N} = h^2 m_X / 16 \pi$, see Eq.~(\ref{Xdec}). We have used $f^{\rm eq}_\psi = f^{\rm eq}_X / f^{\rm eq}_N$, with $f^{\rm eq}_X ({\vec p}_X) = e^{- \beta E_X}$ and $f^{\rm eq}_N ({\vec p}_N) = e^{- \beta p_N}$. 

We note that $p_X dp_X = E_X dE_X$, and hence:
\begin{equation} \label{EE4}
{df_N({\vec p}_N) \over dt} = {m_X\Gamma_{X \rightarrow N} \over p^2_N} (1 -  e^{\beta p_N} f_N({\vec p}_N)) \int_{E_{X,{\rm min}}}^{\infty}{e^{- \beta E_X} d E_X},
\end{equation}
Here, $E_{X,{\rm min}}$ is the minimum value of $E_X$ that results in a given value $p_N$ due to $X$ decay. Conservation of energy and momentum together, see Eq.~(\ref{mom2}), imply:
\begin{equation}
E_X = p_N + \sqrt{p^2_X + p^2_N - 2 p_X p_N {\rm cos} \theta}.
\end{equation}
This gives:
\begin{equation}
{dE_X \over d\theta} = {p^2_X {\rm sin} \theta \over E_X {\rm cos} \theta  - p_X}.
\end{equation}
Thus, the minimum of $E_X$ occurs when $\theta = 0$ or $\theta = \pi$. The former corresponds to forward production of $N$ and is the case when $p_N > m_X/2$. The latter corresponds to backward production of $N$, which happens to be the case when $p_N < m_X/2$. In both cases, we find:
\begin{equation} \label{Emin}
E_{X,{\rm min}} = p_N + {m^2_X \over 4 p_N} .
\end{equation}
We see that, as expected, $E_{X,{\rm min}} \geq m_X$. The lower bound is saturated when $p_N = m_X/2$, which occurs when the decaying $X$ is at rest.

With the final integration, Eq.~(\ref{EE4}) then takes the following form:
\begin{equation} \label{EE5}
{df_N({\vec p}_N) \over dt} = {m_X\Gamma_{X \rightarrow N} T \over p^2_N} (e^{-p_N/T} -  f_N({\vec p}_N)) e^{-(m^2_X/4 p_N T)} ,
\end{equation}
Choosing an initial time $t_{\rm i}$ when $T = T_{\rm i}$, we have $p_N(t) = p_N(t_{\rm i}) a(t_{\rm i})/a(t)$ and $T(t) = T_{\rm i} a(t_{\rm i})/a(t)$, where $a$ is the scale factor of Universe with $a(t) \propto t^{1/2}$ during RD. For simplicity, we neglect any changes in $g_*(T)$ due to the high temperatures we are considering here. Accounting for expansion of the Universe, we thus have:
\begin{equation} \label{EE6}
{df_N({\vec p}_N) \over d{\tilde t}} = {\Gamma_{X \rightarrow N} t_{\rm i} {\tilde t}^{1/2} \over \alpha^2{\tilde p}^2_N} (e^{-{\tilde p}_N} -  f_N({\vec p}_N)) e^{-({\tilde t}/4 {\tilde p}_N)} ,
\end{equation}
where ${\tilde p}_N \equiv p_N(t_{\rm i})/T_{\rm i}$, $\alpha \equiv m_X/T_{\rm i}$, and ${\tilde t} \equiv \alpha^2 t/t_{\rm i}$. This can be rewritten as:
\begin{equation}
{df_N({\vec p}_N) \over f^{\rm eq}_N({\vec p}_N) -  f_N({\vec p}_N)} = {\Gamma_{X \rightarrow N} t_{\rm i} {\tilde t}^{1/2} \over \alpha^2{\tilde p}^2_N} e^{-({\tilde t}/4 {\tilde p}_N)} d{\tilde t} ,
\end{equation}
where $f^{\rm eq}_N({\vec p}_N) = e^{-{\tilde p}_N}$ has no dependence on time. This allows us to perform the integral of both sides exactly. Starting with $f_N({\vec p}_N) = 0$ at $t = t_{\rm i}$ and $H_{\rm i} = 1/2t_{\rm i}$, we then find:
\begin{equation} \label{fN}
f_N({\vec p}_N) = f^{\rm eq}_N({\vec p}_N) \left[1 - {\rm exp} \left(-{\gamma \over 2{\tilde p}^2_N} \int_{\alpha^2}^{m^2_X/T^2}{{\tilde t}^{1/2} e^{-({\tilde t}/4 {\tilde p}_N)} d{\tilde t}} \right) \right].   
\end{equation}
where $\gamma \equiv \Gamma_{X \rightarrow N}/H(T=m_X)$ and we have made use of $H(T=m_X)/H_{\rm i} = \alpha^2$. 

We can now calculate the comoving energy density of $N$ as a function of time, $\rho_N^{\rm co} (t)$, as follows:
\begin{equation}
\rho_N^{\rm co} (t) = {2a^4(t) \over (2 \pi)^3} \int{f_N({\vec p}_N) p_N d^3 p_N} = {2a^4(t) \over 2 \pi^2} \int{f_N({\vec p}_N) p^3_N d p_N} ,
\end{equation}
where the factor of 2 counts the internal degrees of freedom of $N$. Substituting the expression in Eq.~(\ref{fN}) for $f_N({\vec p}_N)$, and noting that $p_N \propto a^{-1}$, we arrive at the following relation:
\begin{equation} \label{co}
{\rho_N^{\rm co} (T) \over \rho^{\rm eq,co}_N(T)} = {\int{f^{\rm eq}_N({\vec p}_N) \left[1 - {\rm exp} \left(-{\gamma \over 2{\tilde p}^2_N} \int_{\alpha^2}^{m^2_X/T^2}{{\tilde t}^{1/2} e^{-({\tilde t}/4 {\tilde p}_N)} d{\tilde t}} \right) \right] {\tilde p}^3_N d{\tilde p}_N} \over \int{f^{\rm eq}_N({\vec p}_N) {\tilde p}^3_N d{\tilde p}_N}}.
\end{equation}
We note that the ratio of the comoving energy density of $N$ to its equilibrium value is mainly sensitive to $\gamma$. Dependence on $\alpha$ only shows up through the lower limit of the integral over ${\tilde t}$ that is related to the initial time.

\begin{figure}[ht!]
    \centering
    \includegraphics[width=0.49\textwidth, trim = .6cm 0cm 1cm 0cm, clip = true]{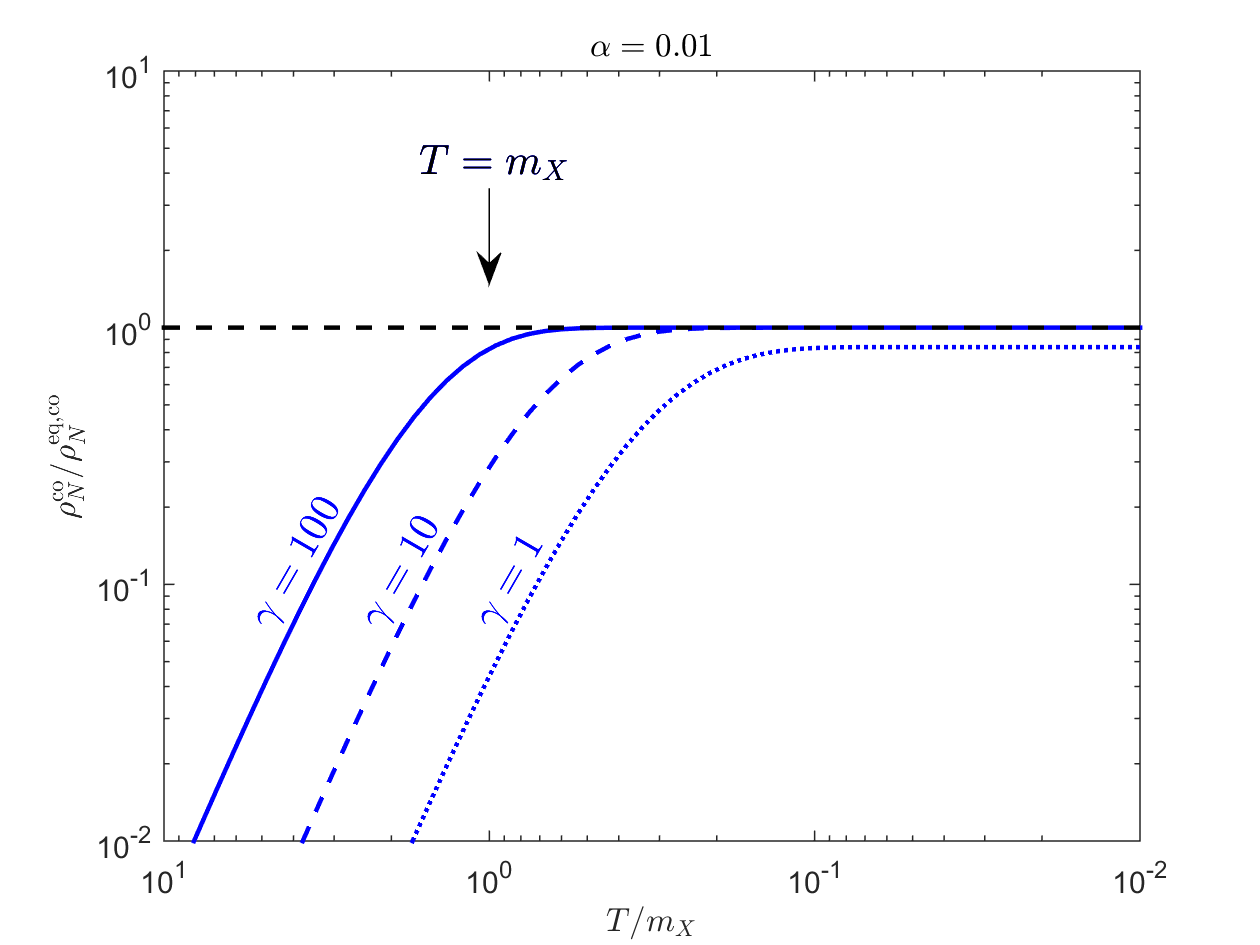}
    \includegraphics[width=0.49\textwidth, trim = .6cm 0cm 1cm 0cm, clip = true]{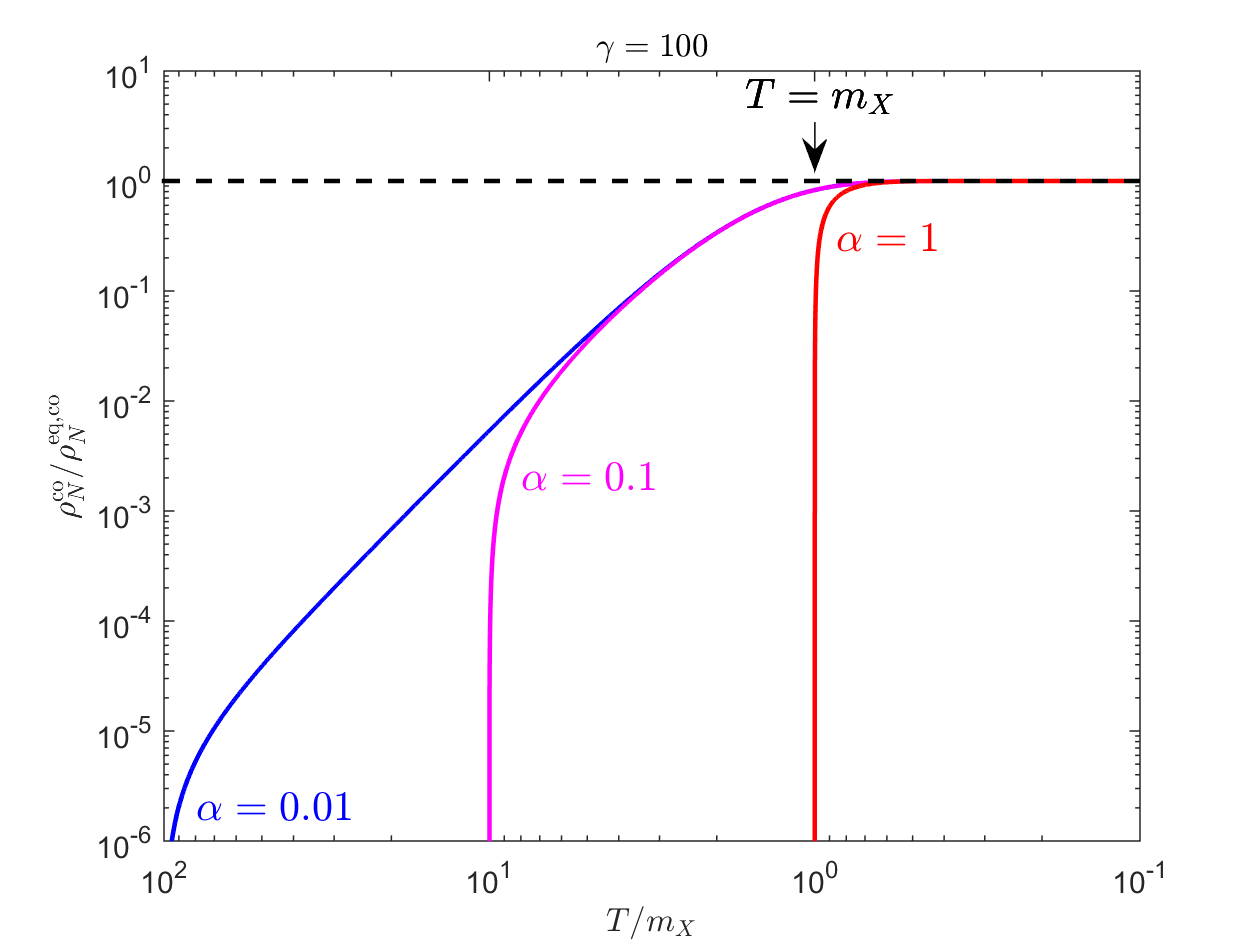}
    \caption{\normalsize Evolution of the comoving energy density of \(N\), normalized to the equilibrium energy density, with respect to temperature, obtained from Eq.~(\ref{co}). In the left panel we vary the parameter \(\gamma\) 
    while in the right panel we vary \(\alpha\). 
    In all cases, the comoving energy density reaches the equilibrium value, shown as the horizontal black dashed line, (or very nearly so for \(\gamma = 1\)) before the temperature drops to \(m_X/10\). From the right panel, we see that 
    there is no significant dependence on the initial time as long as \(T_{\rm i} \gtrsim m_X\). 
    }
    \label{fig:rho}
\end{figure}

In Fig.~\ref{fig:rho} we show the evolution of Eq.~(\ref{co}) with respect to the temperature of the Universe. The horizontal axis corresponds to $T/m_X$ because Eq.~(\ref{co}) depends on this ratio rather than $T$ itself. In the left panel, we see that even for the smallest value allowed in Eq.~(\ref{equi}), expressed as 
\(\gamma = 1\), the comoving energy density of \(N\) reaches \(\sim 0.8 \rho^{\rm eq,co}_N\) around the time when \(T \sim m_X/5\). Thus in our scenario \(N\) acquires a thermal energy density at $T \gg m_N$ as long as \(m_N \lesssim m_X/5\).
However, this would not be the case for $\gamma \ll 1$. The reason being that in this case the maximum comoving energy density established by $X$ decay (and inverse decay) is \(\ll \rho^{\rm eq,co}_N\) as the $X$ number density is Boltzmann suppressed. 
In the right panel, we see that the value of $T/m_X$ for which $\rho^{\rm co}_N \simeq \rho^{\rm eq,co}_N$
is essentially insensitive to $\alpha$, as pointed out above. The only effect of $\alpha$ is to determine the starting point of \(N\) production in our calculation. 

In the above derivation, we have only considered $N$ production from the decay of $X$ (and the accompanying inverse decay). At temperatures $T \gg m_X$, 
inelastic scattering of $N$ off SM particles $N \psi \leftrightarrow \psi^* \psi^*$ might be in equilibrium too. This would be an additional source of $N$ production from the thermal bath and thereby help $\rho^{\rm co}_N$ reach its equilibrium value even faster. Elastic scattering of $N$ off SM particles $N \psi \rightarrow N \psi$ would only redistribute the energy among $N$ particles, which could affect the evolution of individual $f_N({\vec p}_N)$, but would not change the total 
energy density. 

\subsection{Temperature Dependence of Degrees of Freedom } 

\begin{figure}[ht!]
    \centering
    \includegraphics[width=0.49\textwidth, trim = .6cm 0cm 1cm 0cm, clip = true]{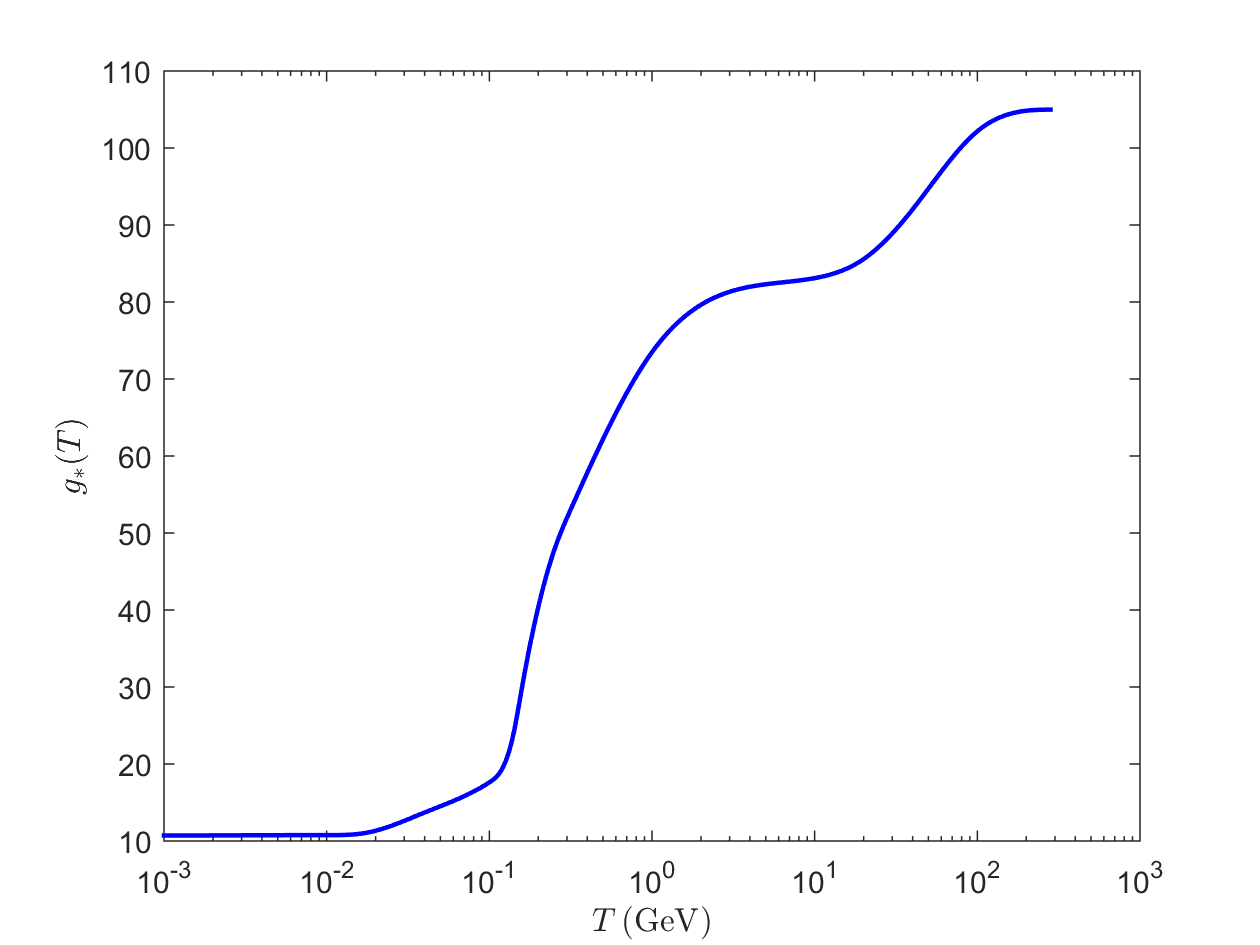}
    \caption{\normalsize Evolution of \(g_*(T)\) as a function of temperature taken from \cite{Borsanyi_gstar}. }
    \label{fig:gstar}
\end{figure}

To compute the number of relativistic degrees of freedom at a given temperature, \(g_*(T)\), in our scenario, we have made use of a smooth continuous function describing the temperature dependence. We do so by utilizing the data presented in Table S2 of \cite{Borsanyi_gstar} with cubic spline interpolation. The resulting curve is shown in Fig.~\ref{fig:gstar}. In our calculations, for temperatures larger than \(\sim 280\,{\rm GeV}\), we use the maximum value of \(g_*(T)\), which is slightly less than the usual 106.75. Of note is the steep decline and abrupt change in slope near \(T \approx 100\) MeV related to the QCD phase transition. This feature is responsible for the larger deviations from linearity seen in our main figures. 

\subsection{Relating the \(h - h'\) and \(m_N - \tau_N\) Planes} 

In Fig.~\ref{fig:mN_tauN} we showed the four conditions labeled 1, 2, 3, and 6, however conditions 4 and 5 of Fig.~\ref{fig:h_hprime} were not shown. Here we present a short discussion to better understand how conditions 4 and 5 affect the interplay between the \(h - h'\) and \(m_N - \tau_N\) planes, and to aid in interpreting Figs.~\ref{fig:h_hprime} and \ref{fig:mN_tauN} together. 

\begin{figure}[ht!]
    \centering
    \includegraphics[width=0.49\textwidth, trim = .6cm 0cm 1cm 0cm, clip = true]{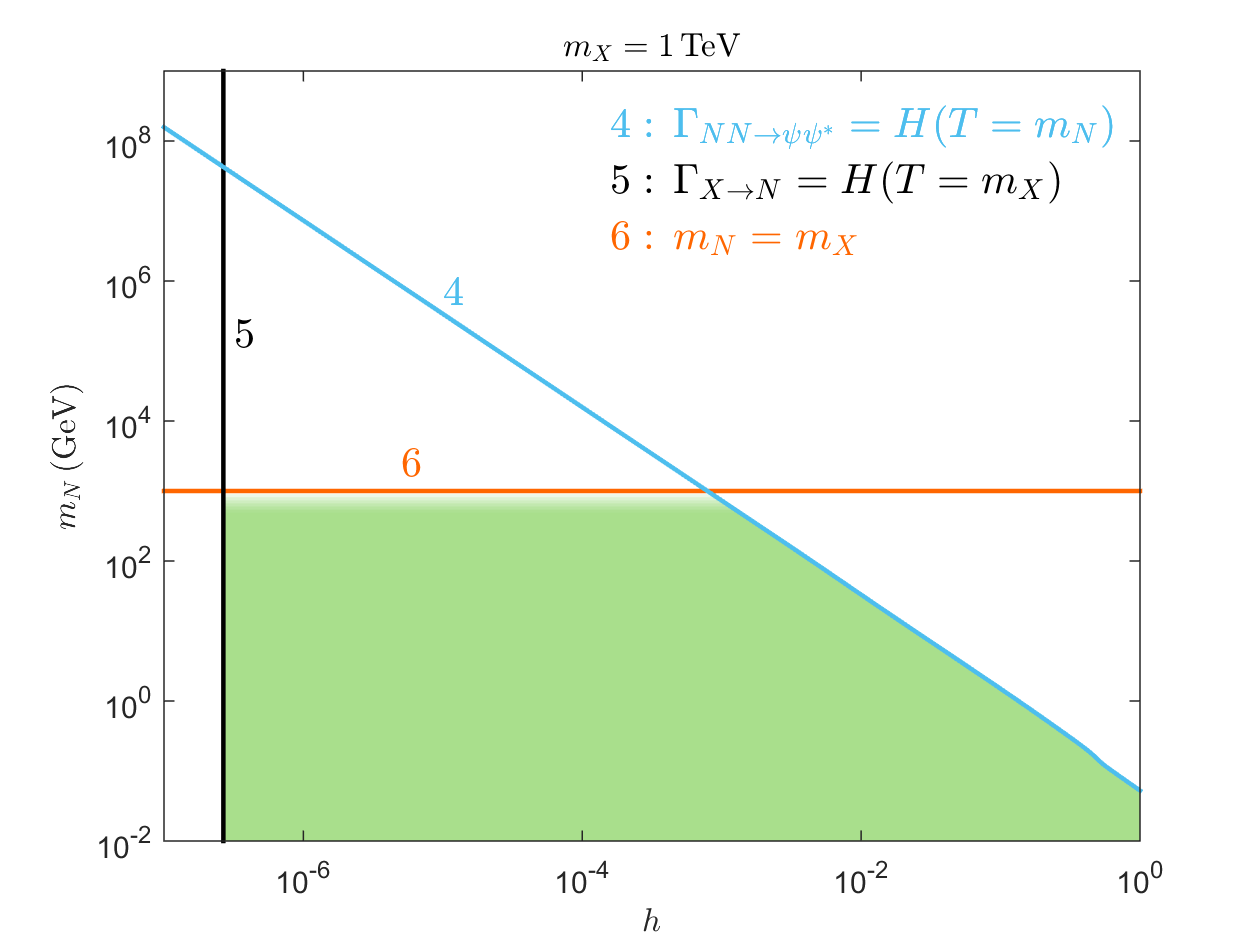}
    \includegraphics[width=0.49\textwidth, trim = .6cm 0cm 1cm 0cm, clip = true]{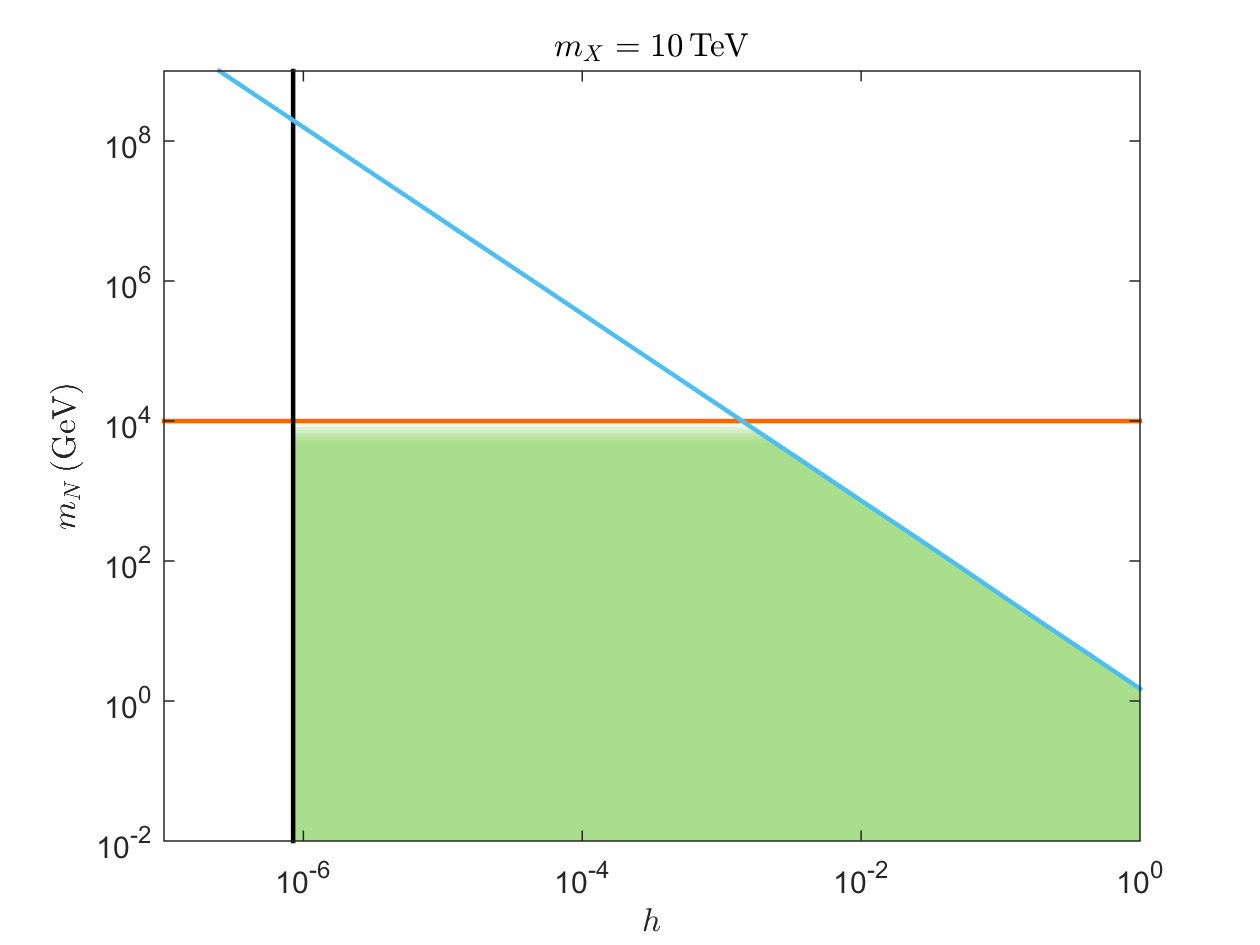}
    \caption{\normalsize Conditions 4, 5, and 6 in the \(m_N - h\) plane for fixed \(m_X\). Lines are colored and labeled as in Figs.~\ref{fig:h_hprime} and \ref{fig:mN_tauN}. The green shaded area is the region allowed by Eqs.~(\ref{equi}) and (\ref{selfann}), as well as \(m_N < m_X\). }
    \label{fig:mN_h}
\end{figure}

We define \(\gamma \equiv \Gamma_{X \rightarrow N}/H(T = m_X)\), as before, and \(\eta \equiv \Gamma_{NN \rightarrow \psi\psi^*}/H(T = m_N)\) such that \(\eta = 1\) and \(\gamma = 1\) correspond to conditions 4 and 5 of Fig.~\ref{fig:h_hprime} respectively. 
The main issue is that a single value of \(m_N\) on Fig.~\ref{fig:mN_tauN} corresponds to a range of \(h\), \(\gamma\), and \(\eta\). Fixing \(m_X = 10\) TeV and \(m_N = 1\) TeV as an example, we can obtain the full range of \(h\), \(\gamma\), and \(\eta\) that are allowed for this single combination of the masses while satisfying both Eqs.~(\ref{equi},\ref{selfann}).
If we choose \(\gamma = 1\) such that we sit on the line for condition 5 in the bottom-right panel of Fig.~\ref{fig:h_hprime}, we get \(h \approx 10^{-6}\) which then gives \(\eta \approx 10^{-16}\). On the other hand, if we first choose \(\eta = 1\) so that we are on the line for condition 4, we then have \(h \approx 10^{-2}\) and therefore \(\gamma \approx 10^8\). The possible range of these three parameters for this single combination of \(m_X\) and \(m_N\) is therefore \(h \approx (10^{-6} - 10^{-2})\), \(\gamma \approx (1 - 10^8)\), and \(\eta \approx (10^{-16} - 1)\). In Fig.~\ref{fig:mN_h} we show conditions 4 and 5, as well as \(m_N = m_X\), in the \(m_N - h\) plane with the corresponding allowed region. Note that \(m_N\) is restricted from below by the other conditions of our scenario, not shown here. 
For a given value of \(m_N\) in Fig.~\ref{fig:mN_h}, one can read off the allowed range of \(h\) and then calculate the corresponding values of the parameters \(\gamma\) and \(\eta\) using Eqs.~(\ref{Xdec},\ref{Nself}). 

We further note that conditions 4 and 5 meet at high \(m_N\), allowing \(\gamma = \eta = 1\) to be simultaneously satisfied. However, this occurs far above \(m_N = m_X\) and is therefore out of reach unless \(m_X\) is itself very large. Additionally, this intersection corresponds to a single value of \(h\) such that the entire allowed region of Fig.~\ref{fig:h_hprime} becomes compressed to a line, as condition 4 moves down to join condition 5. 

\end{document}